\documentclass[11pt]{article}
\pdfoutput=1
\usepackage{graphicx}
\usepackage{latexsym,epsfig,euscript,amssymb,amsmath,verbatim,color, multirow}
\newcommand{\beq}{\begin{equation}}
\newcommand{\eeq}{\end{equation}}
\newcommand{\beqs}{\begin{eqnarray}}
\newcommand{\eeqs}{\end{eqnarray}}
\newcommand{\bit}{\begin{itemize}}
\newcommand{\eit}{\end{itemize}}
\newcommand{\bce}{\begin{center}}
\newcommand{\ece}{\end{center}}
\newcommand{\ben}{\begin{enumerate}}
\newcommand{\een}{\end{enumerate}}
\newcommand{\tr}{\mathrm{tr}}

\newcommand{\nn}{\nonumber}
\newcommand{\MET}{\hbox{$\not  \!\! E_T$ }}
\newcommand{\GHC}{G_{\mathrm{HC}}}

\newcommand{\NHC}{N_{\mathrm{HC}}}

\newcommand{\sw}{{\mathrm s}_{\mathrm w}}
\newcommand{\cw}{{\mathrm c}_{\mathrm w}}
\newcommand{\sww}{{\mathrm s}_{\mathrm {2w}}}
\newcommand{\cww}{{\mathrm c}_{\mathrm {2w}}}

\newcommand{\cwwww}{{\mathrm c}_{\mathrm {4w}}}
\newcommand{\sz}{{\mathrm s}_{\mathrm \zeta}}
\newcommand{\cz}{{\mathrm c}_{\mathrm \zeta}}
\newcommand{\szz}{{\mathrm s}_{\mathrm {2\zeta}}}
\newcommand{\czz}{{\mathrm c}_{\mathrm {2\zeta}}}

\newcommand{\Rcoset}{SU(5)/SO(5)}
\newcommand{\PRcoset}{SU(4)/Sp(4)}
\newcommand{\Ccoset}{SU(4)\times SU(4)'/SU(4)_D}
\newcommand{\Rcolor}{SU(6)/SO(6)}
\newcommand{\PRcolor}{SU(6)/Sp(6)}
\newcommand{\Ccolor}{SU(3)\times SU(3)'/SU(3)_D}

\newcommand{\Fu}{{\mathbf{F}}}
\newcommand{\An}{{\mathbf{A}}}
\newcommand{\Sy}{{\mathbf{S}}}
\newcommand{\Ad}{{\mathbf{Ad}}}
\newcommand{\Sp}{{\mathbf{Spin}}}

\renewcommand{\arraystretch}{1.}
\begin{document}

\pagestyle{empty}

\begin{center}

{\LARGE{\bf Gauge theories of Partial Compositeness:\\
\smallskip

Scenarios for Run-II of the LHC}}

\vspace{1.8cm}

{\large{Gabriele Ferretti}}

\vspace{1.5cm}

{\it Department of Physics, \\
Chalmers University of Technology, \\
Fysikg{\aa}rden 1, 41296 G\"oteborg, Sweden\\
{\tt ferretti@chalmers.se}}

\vspace{2.cm}

\begin{minipage}[h]{14.0cm}

\bce {\bf \large Abstract} \ece

\medskip

We continue our investigation of gauge theories in which the Higgs boson arises as a pseudo-Nambu-Goldstone boson (pNGB) and top-partners arise as bound states of three hyperfermions. All models have additional pNGBs in their spectrum that should be accessible at LHC. We analyze the patterns of symmetry breaking and present all relevant couplings of the pNGBs with the gauge fields. We discuss how vacuum misalignment and a mass for the pNGBs is generated by a loop-induced potential. Finally, we paint a very broad, qualitative, picture of the kind of experimental signatures these models give rise to, setting the stage for further analysis.
\end{minipage}
\end{center}
\newpage

\setcounter{page}{1} \pagestyle{plain} \renewcommand{\thefootnote}{\arabic{footnote}} \setcounter{footnote}{0}

\section{Introduction}

\subsection{Motivation}

The Higgs mechanism~\cite{BEH} in the Standard Model~\cite{Weinberg:1967tq} (SM) does an excellent job at parameterizing the mass spectrum of elementary particles in a consistent way, but leaves many  questions unanswered. We would like to understand why the Higgs mass is so low
and to explain the huge disparity among fermion masses.

One possible explanation of the lightness of the Higgs boson is to realize it as a (pseudo) Nambu-Goldstone Boson (pNGB) of a broken global symmetry. This approach was pioneered in \cite{Kaplan:1983fs} and goes under the name of ``Composite Higgs''.
One way to deal with the disparity of fermionic masses and, in particular, to explain the origin of the top quark mass without reintroducing fine-tuning is to also have additional ``partners'' mixing with SM fermions.
This new ingredient was introduced in \cite{Kaplan:1991dc} and goes under the name of ``Partial Compositeness''.

Much work has been done in this area using the effective field theory description based on the CCWZ formalism~\cite{CCWZ}. There was also a huge effort to realize these construction using extra-dimensions. There are by now exhaustive reviews such as \cite{reviews} providing all the necessary background to these subjects.

A much less studied approach is that of constructing UV completions for these models using a strongly coupled ``hypercolor'' gauge theory with purely fermionic matter (``hyperquarks''). The philosophy behind this proposal is so old fashioned that it almost appears new! Fermionic models of BSM go all the way back to the old technicolor idea and were also tried in the context of composite Higgs and partial compositeness. The recent model building activities try to combine the two. Few explicit proposals have been made so far: \cite{Barnard:2013zea}, \cite{Ferretti:2014qta} and \cite{Vecchi:2015fma} and a partial classification of the available options was made in \cite{Ferretti:2013kya}. (For earlier attempts using supersymmetry, see~\cite{SUSYcomp}.
Alternative avenues being explored are found in~\cite{otherave}.)

The LHC is now entering a phase where the potential for discovery is at its highest point, due to the increase in luminosity and energy. It is thus timely to chart the various scenarios implied by the above class models. In this work we are particularly interested in presenting the underlying theories in detail and in identifying the broad features that may allow one to discern one class of models from the others.
We leave instead a detailed phenomenological analysis for future work. For recent phenomenological work in the area a surely incomplete list is~\cite{AdditionalPheno}.

\subsection{Overview of the results}

In a nutshell, the models we are considering are based on an asymptotically free gauge theory with simple hypercolor group $\GHC$ and fermionic matter in \emph{two} inequivalent irreducible representations (irreps)\footnote{We work with Weyl fermions and count a complex irrep and its conjugate as one.}. The requirement of two different irreps arises from the need to construct top-partners carrying both color and EW quantum numbers. With the notable exception of a model by L.~Vecchi~\cite{Vecchi:2015fma}, this requires at least two separate irreps; one, generically denoted by $\psi$, carrying EW quantum numbers in addition to hypercolor, the other, $\chi$, carrying ordinary color as well as hypercolor.

At low energies, the theory is expected to confine after having spent a part of the RG evolution in or near the conformal window, somewhat in the spirit of~\cite{Luty:2004ye, Strassler:2003ht}.
This is the main dynamical assumption needed for some of the operators in the theory to develop the large anomalous dimensions required to solve the hierarchy problem. However, contrary to the above-mentioned proposal, here we use fermionic operators  \cite{Kaplan:1991dc} to generate the mass of the top quark, eluding the potential problems with fine-tuning pointed out in~\cite{confo}.

Here we are only interested in the behavior of the theory below the dynamically generated scale $\Lambda$, (expected to be of the order of 10~TeV, to fix the ideas). The conformal behavior occurs above this scale, up to the ``flavor'' scale $\Lambda_{\mathrm{UV}} > 10^4$~TeV. In this range the theory could have additional d.o.f./operators driving the conformal behavior and being ultimately responsible for its ending at the scale $\Lambda$.

Below $\Lambda$, the strong IR dynamics of one of the two types of hyperquarks ($\psi$) induces the symmetry breaking needed to realize the composite Higgs scenario.
The three minimal cosets preserving custodial symmetry are $\Rcoset$, $\PRcoset$, and $\Ccoset$. The SM EW group is embedded into the unbroken symmetry. The vacuum is misaligned, inducing a Higgs v.e.v., by the combined action of the one loop potential induced by the SM gauge bosons and the top quark as well as possible hyperquark bare masses of UV origin.

The second irrep ($\chi$)  is needed to realize the QCD color group. Its dynamics may or may not lead to additional pNGBs~\footnote{Note that the condensate $\langle\psi\chi\rangle$ would break the hypercolor group and cannot arise in vector-like theories such as these~\cite{VafaWitten}.}. Top partners arise as $\GHC$ invariant trilinear combinations of the two types of hyperquarks. The top quark acquires a mass via a linear coupling of these partners to the SM fields $Q_L^3 \equiv (t_L, b_L)$ and $u^3_R \equiv t_R$. The remaining SM fields may instead be coupled bilinearly and acquire a mass via the more standard mechanism. This hybrid solution, proposed in~\cite{Ferretti:2014qta, Matsedonskyi:2014iha, Cacciapaglia:2015dsa}, has the extra advantage of suppressing unwanted contributions to dipole moments or flavor violating operators  and could be realized at low energies via the mechanism explained in~\cite{Panico:2016ull}.

With the exception of the Wess-Zumino-Witten (WZW) term, we consider only SM tree level couplings that preserve a parity symmetry, $P_\pi$, changing sign to all the pNGBs except for the Higgs itself. Heavier pNGBs thus decay into lighter ones plus a SM gauge boson or a pair of SM fermions if the decay into a gauge boson is not kinematically allowed.

This parity symmetry is however broken in some cases by the anomaly encoded in the WZW term, and this allows the lightest pGNBs to decay via di-bosons with a very narrow, but still prompt, decay width. It is interesting to notice that~\cite{Ma:2015gra}, for the coset $\Ccoset$, the decay of some of the pNGBs is forbidden by the existence of another symmetry, $G_\pi$, thus providing a possible Dark Matter candidate. For the scope of this paper we only assume that in the $\Ccoset$ scenario the lightest pNGB odd under this additional symmetry is collider stable, leading to the usual signatures--\MET or
highly ionizing tracks depending on the charge. (The requirement of this pNGB being neutral is necessary only in order to have a DM candidate, not simply a collider stable particle.)

The leading production mode for the pNGBs associated with the EW coset are Drell-Yan production and vector boson fusion.

If the dynamic in the color sector also leads to symmetry breaking, (as we assume through the paper for illustration purposes, since this case leads to additional interesting phenomena), there will be additional colored pNGB with a mass higher than the EW ones since it is due to gluon loops. All models have a neutral pNGB in the octet of color that can be singly produced and decay via an anomalous coupling. Some models also include additional charged and colored pNBGs in the triplet or sextet that, under the assumption of $P_\pi$-parity, decay to two jets and a lighter EW pNGB. Their charges are fixed by the structure of the top partners.

An universal feature of all of these models is the presence of two additional scalars arising from the two spontaneously broken $U(1)$ axial symmetries associated to the two fermionic irreps. One of these bosons is associated to a $\GHC$ anomalous current and it is thus expected to acquire a large mass just like the $\eta'$ in QCD. The remaining one is instead naturally light in the absence of additional UV mechanisms such as bare hyperquark masses. Both couple to gluons via the anomaly and could provide an explanation of the current 750~GeV di-photon excess~\cite{diphotonATLASCMS}. Indeed, such an interpretation has already been put forward in~\cite{Belyaev:2015hgo} for the case of the light $U(1)$ boson. (More details about the role of pNGBs in explaining the excess are given in~\cite{Bellazzini:2015nxw}.)

\subsection{Organization of the paper}

The paper is organized as follows:
In Section~2 we present the class of models of interest. We then turn to study their different sectors beginning in Section~3 with the pNGBs associated to the EW coset. We study the generation of the potential, its symmetries, present a couple of prototypical spectra, work out all the couplings of relevance for LHC physics and briefly comment on the main phenomenological aspects. In Section~4 we discuss the colored objects in the different theories, pNGBs and top partners, show how their quantum numbers are related and how this affects the phenomenology. In Section~5 we comment on the remaining two pNGBs universally present in this class of models.

Technical details are collected in the appendix. Appendix A lists all the gauge theories having a composite higgs and a top partner under the requirements discussed in Section~2 and~3 and discusses their IR properties. Appendix B contains the conventions for the explicit construction of the EW cosets. Appendix C lists additional couplings (anomalous and non) that did not find a place in the main text.

\section{The models, streamlined classification}

In this section we summarize the models of interest in this paper. We take the opportunity to slightly expand and streamline the classification presented in \cite{Ferretti:2013kya}.

We want to realize the ``composite Higgs'' coset by condensation of a set of fermionic hyperquarks $\psi$ transforming in some irrep of a simple hypercolor gauge group $\GHC$.
Recall that the three  basic cosets one can realize with fermionic matter depend on the type of irrep to which the fermions belong.
One possibility is to mimic ordinary QCD. Working with left-handed (LH) fermions only, a set of $n$ \emph{pairs} of LH fermions $(\psi_i, \tilde \psi^i)$ in a $(R, \bar R)$ irrep of $\GHC$, with $R$ complex (C) and $\bar R$ its conjugate,
breaks the global symmetry $SU(n)\times SU(n)'\to SU(n)_D$ after condensation $\langle \tilde\psi^i \psi_j\rangle \propto \delta^i_j$.
(The $ U(1)$ factors will be studied separately because of possible ABJ anomalies. Here we concentrate on the non-abelian factors.)

If, on the other hand, we consider just a single set of $n$ LH fermions $\psi_i$ in a real (R) (respectively pseudo-real (PR)) irrep, the symmetry breaking is
$SU(n)\to SO(n)$ (resp. $SU(n)\to Sp(n)$) since the condensate $\langle\psi_i \psi_j\rangle$ turns out to be symmetric (resp. anti-symmetric).

If we want to use such cosets to construct an EW sector for the composite Higgs, the possible minimal custodial cosets of this type are
$SU(4)\times SU(4)'/SU(4)_D$, $SU(5)/SO(5)$ and $SU(4)/Sp(4)$ for the three cases. For instance, $SU(4)/SO(4)$ is not acceptable since the pNGBs are only in the symmetric irrep $(\mathbf{3},\mathbf{3})$ of $SO(4)=SU(2)_L\times SU(2)_R$ and thus we do not get the Higgs irrep $(\mathbf{2},\mathbf{2})$.

Since we want to obtain the top partners as fermionic trilinears, we also need to embed the color group $SU(3)_c$ into the global symmetry of the composite theory.
For this purpose we introduce a second fermionic irrep $\chi$ coupling to color as well as hypercolor.
The minimal field content allowing an anomaly-free embedding of unbroken $SU(3)_c$ are $SU(3)\times SU(3)'\rightarrow SU(3)_D \equiv SU(3)_c$ for the complex case, $SU(6) \rightarrow SO(6) \supset SU(3)_c$ for the real case and $SU(6)\rightarrow Sp(6) \supset SU(3)_c$ for the pseudoreal case.

In all of these cases we need 6 LH fermions altogether, to be divided into three pairs $(\chi, \tilde\chi)$ in the case of a complex irrep.
Top-partners are constructed by $\GHC$ invariant trilinears of type $\psi\chi\psi$ or $\chi\psi\chi$ depending on the model as shown in Appendix~A.

All combinations of R, PR and C irreps are in principle possible.
The minimal cosets are shown in  Table~\ref{tablecos}. The three cases crossed out are those that do not give rise to top partners. This can be easily seen e.g. for the case in which both irreps are pseudo-real since the product of three pseudo-real irreps cannot contain a singlet. For each remaining case one can look for possible hypercolor gauge groups and irreps that satisfy the remaining constraint of asymptotic freedom. These are listed in Appendix~A for completeness. More details can be found in~\cite{Ferretti:2013kya}.

\begin{table}
\begin{picture}(100,100)
\thicklines
\put(40,90){{\bgroup
\def\arraystretch{2.2}%

\begin{tabular}{|c|c|c|c|}
 \multicolumn{1}{c}{} &  \multicolumn{1}{c}{$\psi\in$ R} &  \multicolumn{1}{c}{$\psi\in$ PR} &  \multicolumn{1}{c}{ $\psi, \tilde \psi \in$ C}\\
  \cline{2-4}
    \multicolumn{1}{c|}{\!\!\!\!\!\!\!\!\!\!\!\!\!\!\!\!\! $\chi\in$ R }  &  ${ \frac{SU(5)}{SO(5)}}{ \frac{SU(6)}{SO(6)}} U(1)_u$&
     ${ \frac{SU(4)}{Sp(4)}}{ \frac{SU(6)}{SO(6)}} U(1)_u$ & ${ \frac{SU(4)\times SU(4)'}{SU(4)_D}}{ \frac{SU(6)}{SO(6)}} U(1)_u$ \\
  \cline{2-4}
     \multicolumn{1}{c|}{\!\!\!\!\!\!\!\!\!\!\!\!\!\!\!\!\!\!\!\!\!\! $\chi\in$  PR } &  ${ \frac{SU(5)}{SO(5)}}{ \frac{SU(6)}{Sp(6)}} U(1)_u$&
       ${ \frac{SU(4)}{Sp(4)}}{ \frac{SU(6)}{Sp(6)}} U(1)_u$ & ${ \frac{SU(4)\times SU(4)'}{SU(4)_D}}{ \frac{SU(6)}{Sp(6)}} U(1)_u$ \\
  \cline{2-4}
     \multicolumn{1}{c|}{\!\!\!\!\!\!\!\!\!\!\!\!\!\!\!\!\!\!\!\!\!\!\!\! $\chi, \tilde\chi \in$ C }  & ${ \frac{SU(5)}{SO(5)}}{ \frac{SU(3)\times SU(3)'}{SU(3)_D}} U(1)_u$&
       ${ \frac{SU(4)}{Sp(4)}}{ \frac{SU(3)\times SU(3)'}{SU(3)_D}} U(1)_u$ \!\!\!\!&
      ${ \frac{SU(4)\times SU(4)'}{SU(4)_D}}{ \frac{SU(3)\times SU(3)'}{SU(3)_D}} U(1)_u$\\
  \cline{2-4}
\end{tabular}
}
\egroup}
\put(180,60){ \line(4,1){90}}
\put(180,80){ \line(4,-1){80}}
\put(190,20){ \line(4,1){100}}
\put(185,42){ \line(4,-1){90}}
\put(320,60){ \line(4,1){90}}
\put(320,85){ \line(4,-1){100}}
\end{picture}
\caption{The possible minimal cosets realized in this class of models. The hyperquarks $\psi$ and $\chi$ transform under different irreps of $\GHC$. $\psi$ also carries EW quantum numbers, while $\chi$ carries color. The three cases crossed out are those that do not give rise to top partners because the nature of their congruency classes prevents the formation of singlets.}
  \label{tablecos}
\end{table}

Table~\ref{tablecos} also shows a ``ubiquitous'' non-anomalous $U(1)_u$ factor arsing from the spontaneous breaking of the $\GHC$-anomaly-free abelian chiral symmetry.
This symmetry is obtained by constructing the anomaly free linear combination of the two axial symmetries $U(1)_A^\psi$ and $U(1)_A^\chi$ rotating, respectively, all the $\psi$ (or $\psi, \tilde\psi$) and $\chi$ (or $\chi, \tilde\chi)$ by the same phase.
For each pair of complex irreps there is also one vector-like $U(1)_V^\psi$ or $U(1)_V^\chi$ factor which is both anomaly free and unbroken.

To understand the type of pNGBs arising in the various cases, we look at the decomposition under $SU(2)_L\times U(1)_Y$ of the irrep of $H$ under which the pNGB's transform\footnote{We denote specific irreps either by their dimensionality or by the symbols $\Fu,~\Sy_n,~\An_n,~\Ad$ and $\Sp$ for the fundamental, $n-$symmetric, $n-$antisymmetric, adjoint and spin.}. The decomposition is shown in Table~\ref{decopNGB}.
\medskip

\begin{table}
\bce
{\footnotesize
\begin{tabular}{|c|c|c|}
\hline
$G/H$ & irrep of $H$ & $H\to SU(2)_L\times U(1)_Y$\\
\hline\hline
$\Rcoset$ & ${\mathbf{14}}\equiv \Sy_2$ & ${\mathbf 3}_{\pm 1}(\phi_\pm) + {\mathbf 3}_0(\phi_0) + {\mathbf 2}_{\pm 1/2}(H) + {\mathbf 1}_0(\eta)$\\
$\PRcoset$ & ${\mathbf{5}} \equiv \An_2$ & $ {\mathbf 2}_{\pm 1/2}(H) + {\mathbf 1}_0(\eta)$ \\
$\Ccoset$ & ${\mathbf{15}}\equiv \Ad$ & $ {\mathbf 3}_0(\phi_0)+ {\mathbf 2}_{\pm 1/2}(H) +  {\mathbf 2'}_{\pm 1/2}(H') + {\mathbf 1}_{\pm
1}(N_\pm) + {\mathbf 1}_0(N_0) + {\mathbf 1'}_0 (\eta)$ \\
\hline
\end{tabular}}
\ece
\caption{The decomposition under $SU(2)_L\times U(1)_Y \subset H$ of the pNGBs associated to the coset $G/H$ and transforming linearly under an irrep of $H$.
Next to the charges we indicate the names used throughout the paper. For uniformity of notation we denote all $SU(2)_L$ triplets by $\phi$ and doublets by $H$. The fields $N$ in $\Ccoset$ form a triplet of $SU(2)_R$.}
  \label{decopNGB}
\end{table}

The EW cosets above have been studied in many previous papers, see e.g. \cite{earlySp, Katz:2005au,Lodone:2008yy,Gripaios:2009pe, Cacciapaglia:2014uja} for $\PRcoset$, \cite{Schmaltz:2010ac, Ma:2015gra} for $\Ccoset$ and \cite{Georgi:1984af, Dugan:1984hq, ArkaniHamed:2002qy, Katz:2005au, Vecchi:2013bja, Ferretti:2014qta} for $\Rcoset$. General non-minimal cosets are discussed in~\cite{Mrazek:2011iu}.

As for the color cosets, arising when the $\chi$ also condense, a generic prediction is the existence of an electrically neutral color octet pNGB. In addition, we have a pair of electrically charged pNGBs in the $({\mathbf{3}}, \overline{\mathbf{3}})$ of $SU(3)_c$ for the $SU(6)/Sp(6)$ case or in the  $({\mathbf{6}}, \overline{\mathbf{6}})$ for the $SU(6)/SO(6)$ case. The charges are discussed in Section~4.

Top partners can be broadly divided into two separate groups: those of type $\psi\chi\psi$ and those of type $\chi\psi\chi$. (We are being schematic here, and only indicate the relative number of $\psi$ or $\chi$-type hyperquarks, without indicating the specific Lorentz and hypercolor contractions.) Top partners of the first type require coupling to top quark spurions in a two index irrep, while partners of the second type give rise to single index irreps.

There is a sense in which models of type $\psi\chi\psi$ are more promising than the others. Top-partners of type $\chi\psi\chi$ force one to chose the fundamental irrep for the spurions. For the $SU(5)$ case this leads to the ${\mathbf{5}}$ that, although being compatible with the $Z\to b_L \bar b_L$ custodial symmetry~\cite{Agashe:2006at, Ferretti:2014qta}, gives rise to effective potentials that tend to break the usual custodial symmetry~\cite{Sikivie:1980hm}. The case of $SU(4)\times SU(4)'$ leads to problems already at the $Z\to b_L \bar b_L$ level and we exclude these models from the list in Appendix~A. There are no $\chi\psi\chi$ cases for the $\PRcoset$ coset.

\section{The Electro-Weak sector}
\subsection{The potential}

The pNGBs acquire a mass from a loop-induced~\cite{Coleman:1973jx} potential that breaks the shift symmetry explicitly.
We consider three kinds of contribution to the potential. The first one is the contribution from the loop of gauge bosons, which is uniquely determined by the gauge structure up to an overall dimensionless positive constant $B$.
 It can be written as\footnote{We chose to use the pNGB decay constant $f$ as the only dimensionfull parameter. This simplifies the notation but hides the scaling properties of the formulas. See Appendix B for the conventions on the generators and the non-linear pNGB matrix $U$.}
\beq
    V_g = B\; f^4 \; \tr(g^2 T_L^a U T_L^{a *} U^\dagger + g'^2 T_R^3 U T_R^{3 *} U^\dagger)
\eeq
for the $SU(4)/Sp(4)$ case and
\beq
    V_g = -B\; f^4 \; \tr(g^2 T_L^a U T_L^a U^\dagger + g'^2 T_R^3 U T_R^3 U^\dagger)
\eeq
for $SU(4)\times SU(4)'/SU(4)_D$. For the $SU(5)/SO(5)$ coset both expressions are equivalent in our conventions from Appendix~B. Actually, for all three cases the formula could be written in a uniform notation using the matrix $\Sigma$ defined in Appendix B instead of $U$, but we choose to work with $U$ because of its easier transformation properties under the full symmetry group.

The constant $B$ and related ones are the so-called low-energy-coefficients (LEC) (in units of $f$) that
encode the information about the spectrum of the strongly interacting theory. Lacking direct experimental information, they could be estimated on the lattice. Some work in this direction has already been done in the context of a specific model~\cite{lattice}. (For more general results on the lattice, see the review~\cite{DeGrand:2015zxa}.) These models necessarily involve representations of the hypercolor group other than the fundamental and pose additional challenges. In the context of phenomenology they have also been used in e.g.~\cite{higher}. For a clear discussion of how they are generated and can be computed in the context of effective theories of partial compositeness, we refer to~\cite{Golterman:2015zwa} and references therein.

We also have the option of adding bare hyper-quark masses
\beq
    V_m = -B' \;f^4\; \tr(\mu U + \mu^* U^*)
\eeq
with $\mu$ a dimensionless matrix preserving the custodial symmetry and $B'$ some other dimensionless constant. For definitiveness we take $\mu={\mathbf{1}}$ if needed.

Lastly, we need to take into account the effect of the top quark, which leads to vacuum misalignment~\cite{Agashe:2004rs}. This can be done introducing spurionic fields transforming under a particular irrep of the unbroken flavor group. Here is one instance when having a candidate UV completion helps in picking the particular irreps to consider.

We restrict to the case where only the Higgs acquires a v.e.v. since we want to preserve the SM-like properties of the Higgs boson as well as the tree level mass relation $m_W^2 = \cos^2\theta_W m_Z^2$.
Since we are only allowing the Higgs direction to be turned on, the matrix of v.e.v.s is easily exponentiated and we find it convenient to introduce a matrix $\Omega(\zeta)$ for all three cases denoting the vacuum misalignment and depending on $v=246$~GeV through $\sin\zeta = v/f$. In terms of the original Higgs field $\hat h$ gaining a v.e.v. we have $\zeta = \langle\hat h \rangle/f$. In other words $v = f \sin(\langle\hat h \rangle/f)$. The fields appearing into the effective lagrangian are always the canonically normalized fields with zero v.e.v.. The expression for $\Omega$ and $U$ is found in Appendix~B.

It is then a straightforward matter to check which, among the irreps of $G=SU(5), SU(4), SU(4)\times SU(4)$ with up to two indices, contains spurions for the third generation quarks that preserve the custodial symmetry in the sense specified above. The decomposition of $G\to SU(2)_L\times SU(2)_R$ is useful at this point and it is reproduced in Table~\ref{decompositions} for convenience.

\begin{table}
{\footnotesize
\begin{tabular}{c|rcl}
  $G$ &  &  & $SU(2)_L\times SU(2)_R$ \\
  \hline
 $SU(5)$  & ${\mathbf{1}}$ & $\to$ & $(\mathbf{1}, \mathbf{1})$ \\
   & ${\mathbf{5}}, {\overline{\mathbf{5}}}$ & $\to$ & $(\mathbf{1}, \mathbf{1}) + (\mathbf{2}, \mathbf{2}) $ \\
   & ${\mathbf{10}}, {\overline{\mathbf{10}}}$ & $\to$ & $(\mathbf{2}, \mathbf{2}) + (\mathbf{1}, \mathbf{3}) + (\mathbf{3}, \mathbf{1}) $ \\
   & ${\mathbf{15}}, {\overline{\mathbf{15}}}$ & $\to$ & $2\times(\mathbf{1}, \mathbf{1}) + (\mathbf{2}, \mathbf{2}) + (\mathbf{3}, \mathbf{3}) $ \\
   & ${\mathbf{24}}$ & $\to$ & $(\mathbf{1}, \mathbf{1}) + 2\times (\mathbf{2}, \mathbf{2}) + (\mathbf{1}, \mathbf{3}) + (\mathbf{3}, \mathbf{1})
   + (\mathbf{3}, \mathbf{3}) $ \\
  \hline
 $SU(4)$   & ${\mathbf{1}}$ & $\to$ & $(\mathbf{1}, \mathbf{1})$ \\
   & ${\mathbf{4}}, {\overline{\mathbf{4}}}$ & $\to$ & $(\mathbf{1}, \mathbf{2}) + (\mathbf{2}, \mathbf{1})$ \\
   & ${\mathbf{6}}$ & $\to$ & $2 \times (\mathbf{1}, \mathbf{1}) + (\mathbf{2}, \mathbf{2})$ \\
   & ${\mathbf{10}}, {\overline{\mathbf{10}}}$ & $\to$ & $(\mathbf{2}, \mathbf{2}) + (\mathbf{1}, \mathbf{3}) + (\mathbf{3}, \mathbf{1})$ \\
   & ${\mathbf{15}}$ & $\to$ & $(\mathbf{1}, \mathbf{1}) + 2\times (\mathbf{2}, \mathbf{2}) + (\mathbf{1}, \mathbf{3}) + (\mathbf{3}, \mathbf{1})$ \\
  \hline
  $SU(4)\times SU(4)'$  & $(\mathbf{1}, \mathbf{1})$ & $\to$ & $(\mathbf{1}, \mathbf{1})$ \\
   & $(\mathbf{1}, \mathbf{4}), \dots(\overline{\mathbf{4}}, \mathbf{1})$ & $\to$ & $(\mathbf{1}, \mathbf{2}) + (\mathbf{2}, \mathbf{1})$ \\
   & $(\mathbf{1}, \mathbf{6}), (\mathbf{6}, \mathbf{1})$ & $\to$ & $2 \times (\mathbf{1}, \mathbf{1}) + (\mathbf{2}, \mathbf{2})$ \\
   & $(\mathbf{1}, \mathbf{10}), \dots (\overline{\mathbf{10}}, \mathbf{1})$ & $\to$ & $(\mathbf{2}, \mathbf{2}) + (\mathbf{1}, \mathbf{3})
   + (\mathbf{3}, \mathbf{1})$ \\
   & $(\mathbf{1}, \mathbf{15}), (\mathbf{15}, \mathbf{1})$ & $\to$ & $(\mathbf{1}, \mathbf{1}) + 2\times (\mathbf{2}, \mathbf{2})+
    (\mathbf{1}, \mathbf{3}) + (\mathbf{3}, \mathbf{1})$\\
   & $(\mathbf{4}, \mathbf{4}), \dots(\overline{\mathbf{4}}, \overline{\mathbf{4}})$  & $\to$ & $2 \times (\mathbf{1}, \mathbf{1})
     + 2 \times (\mathbf{2}, \mathbf{2})  + (\mathbf{1}, \mathbf{3}) + (\mathbf{3}, \mathbf{1})$ \\
   \hline
\end{tabular}}
\caption{Decompositions of the irreps of $G$ to be used to identify candidate spurions.}
\label{decompositions}
\end{table}

Spurions containing a $(\mathbf{2}, \mathbf{2})$ are possible candidates for $Q^3_L$ and spurions containing $(\mathbf{1}, \mathbf{1})$,  $(\mathbf{1}, \mathbf{2})$ or $(\mathbf{1}, \mathbf{3})$ are candidates for $t_R$. Spurions in the $(\mathbf{2}, \mathbf{1})$ should not be used because they violate the extra custodial requirement~\cite{Agashe:2006at}.

The spurions irrep should be matched with the type of baryon arising in the UV completion. If, in a particular model, the composite top partners arise from bound states of type $\chi\psi\chi$, then the spurions to be used are those in the one index irrep (the fundamental). Vice-versa, if the top partners in a model are of type $\psi\chi\psi$, one should use two indices irreps, to be further restricted to symmetric, anti-symmetric, adjoint or bi-fundamental if required by the symmetries of the particular model. From Table~\ref{allmodels} in Appendix~A one can reconstruct the requirements case by case.

A spurion $S$ in a two-index irrep of $SU(n)$ transforms as $S\to g S g^T$ if in the $\Sy_2$ or $\An_2$ irrep and $S\to g S g^\dagger$ if in the $\Ad$.
In the $\Ccoset$ case one should instead talk about $(\Fu, \Fu)$ or $(\Fu, \overline \Fu)$, whereby $S \to g S g'^T$ or $S \to g S g'^\dagger$.
Similarly (see Appendix~B) the symmetry properties of the pNGB field $U$ are $U \to g U g^T$ for the $\Rcoset$ and $\PRcoset$ cosets and
$U \to g U g'^\dagger$ for $\Ccoset$. Thus, we see that, to leading order, the potential for two-index representations is proportional to the expressions in Table~\ref{spurionpotential}. Spurions like $(\Fu,\overline \Fu)$ must couple to top partners containing one $\psi$ and one $\tilde \psi$. Spurions of the type $(R,{\mathbf{1}})$ or $({\mathbf{1}}, R')$ such as $(\Fu, {\mathbf{1}})$, $(\An_2, {\mathbf{1}})$ etc., do not give rise to a non-trivial invariant since we need to multiply directly $U$ and $U^\dagger$.

\begin{table}
\def\arraystretch{2.}%
\hspace{1cm}{\footnotesize
\bce
\begin{tabular}{|c|c|c|c|}
 \multicolumn{1}{c}{} &  \multicolumn{1}{c}{$ S \in \Sy_2 $} &  \multicolumn{1}{c}{$ S \in \An_2 $} &  \multicolumn{1}{c}{ $ S \in \Ad $}\\
  \cline{2-4}
    \multicolumn{1}{c|}{\!\!\!\!\!\!\!\!\!\!\!\!\!\!\!\!\! $\Rcoset$ }  &  $\tr(S U^*) \tr(S^* U)$&
     $0$ & $ \tr(S U S^* U^*) $ \\
       \cline{2-4}
         \multicolumn{1}{c|}{\!\!\!\!\!\!\!\!\!\!\!\!\!\!\!\!\! $\PRcoset$ }  &  $ 0 $&
     $ \tr(S U^*) \tr(S^* U)$ & $ \tr(S U S^* U^*) $ \\
       \cline{2-4}
\end{tabular}

\begin{tabular}{|c|c|c|}
 \multicolumn{1}{c}{} &  \multicolumn{1}{c}{$ S \in (\Fu,\Fu) $} &  \multicolumn{1}{c}{$ S \in (\Fu,{\overline{\Fu}}) $} \\
  \cline{2-3}
         \multicolumn{1}{c|}{\hspace{-2cm} $\Ccoset$ }  &  $ \tr(U S^T U^* S^\dagger) $&$ \tr(S U^\dagger) \tr(S^\dagger U) $ \\
 \cline{2-3}
\end{tabular}\ece}
\caption{The spurion couplings at leading order for the two index irreps, to be associated to models where the top partners are of type $\psi\chi\psi$. The zeros arise in the case when $U$ and $S$ have opposite symmetry properties. Irreps of type $(R,{\mathbf{1}})$ or $({\mathbf{1}}, R')$ such as $(\Fu, {\mathbf{1}})$, $(\An_2, {\mathbf{1}})$ etc. for $\Ccoset$ do not give rise to a non-trivial invariant since we need to multiply directly $U$ and $U^\dagger$.}
\label{spurionpotential}
\end{table}

In the cases of $\Rcoset$ one could also consider spurions in the fundamental $\Fu$ of $SU(5)$. In this case the leading contribution to the potential is of forth order and proportional to\footnote{We ignore possible non factorizable contributions and refer again to~\cite{Golterman:2015zwa} for details.}
\beq
            (S^\dagger U S^*) (S^T U^* S).
\eeq
The $\Fu$ for the coset $\Rcoset$ runs into trouble with the desire to have a vacuum that preserves custodial symmetry. In this case, coupling generically the pNGBS to spurions in the fundamental will induce a tadpole for the field $\phi_+^- - \phi_-^+$ which should be suppressed in order to avoid tree level corrections to the $\rho$-parameter. If we were to take this fact also as a strict guideline, we would be led to exclude \emph{all} the cases in Appendix~A giving top partners of type $\chi\psi\chi$, although this may be a bit too drastic at this stage.

In the above formulas $S$ could carry a $SU(2)_L$ index in the case it corresponds to $Q^3_L$. This index is then also summed over in the obvious way.
Notice that terms proportional to $\tr(S U^*)+\tr(S^* U)$ or $(S^\dagger U S^*)+ (S^T U^* S)$ are not allowed due to the need to preserve the spurionic $U(1)$.

\subsection{The parity transformations $P_\pi$ and $G_\pi$}

We are now in the position of defining more concretely the parity symmetries of relevance for these models, starting with $P_\pi$. For the scope of this paper we will think of $P_\pi$ as an accidental symmetry of the non-anomalous pNGB Lagrangian coupled to the SM. Its action changes sign to all the pNGB except the Higgs doublet(s) and can be realized in all three cases as $ U \to \hat P_\pi U^\dagger  \hat P_\pi$ with the matrix $ \hat P_\pi$ defined as

{\footnotesize
\beq
   \hat P_\pi = \left(\begin{array}{ccccc}
 1 & 0 & 0 & 0 & 0 \\
 0 & 1 & 0 & 0 & 0 \\
 0 & 0 & 1 & 0 & 0 \\
 0 & 0 & 0 & 1 & 0 \\
 0 & 0 & 0 & 0 & -1
\end{array} \right),
\quad
  \hat  P_\pi =  \left(\begin{array}{cccc}
 1 & 0 & 0 & 0 \\
 0 & 1 & 0 & 0 \\
 0 & 0 & -1 & 0 \\
 0 & 0 & 0 & -1
\end{array} \right),
\quad
 \hat  P_\pi =  \left(\begin{array}{cccc}
 0 & 1 & 0 & 0 \\
 -1 & 0 & 0 & 0 \\
 0 & 0 & 0 & 1 \\
 0 & 0 & -1 & 0
\end{array} \right)
\eeq}

\noindent for the three cosets $\Rcoset$, $\Ccoset$ and $\PRcoset$ respectively.

To see that the transformation accomplishes its task note first that
$ \hat P_\pi \Omega^* = \Omega \hat  P_\pi$ for $\Rcoset$ and $\PRcoset$ and $ \hat P_\pi \Omega^\dagger= \Omega  \hat P_\pi$ for $\Ccoset$. This allows one to move the action of $ \hat P_\pi$ pass the vacuum misalignment matrix directly onto the pNGB matrix $\Pi$ (c.f.r. Appendix B) where its effect is to reverse the sign of the Higgs doublet(s). This, together with the hermitian conjugation on $U$ that reverses the sign of all pNGBs, has the desired combined effect. In all three cases $P_\pi$ leaves the vacuum invariant and
preserves the custodial symmetry group. In particular $D_\mu ( \hat P_\pi U^\dagger  \hat P_\pi) = \hat  P_\pi (D_\mu U) ^\dagger  \hat P_\pi$.

Note that the hermitian conjugation is necessary in all three cases. But it is known that the WZW term breaks precisely this last transformation and thus $P_\pi$ can never be an exact symmetry at the quantum level. Still, it is desirable for the Yukawa couplings to be left invariant by such transformation since this prevents the generation of custodial symmetry breaking v.e.v.s from the induced potential and greatly alleviates the constraints from flavor physics, e.d.m. etc. This condition can be realized by imposing the invariance of the spurion fields. In particular, for the two-index irreps in Table~\ref{spurionpotential} we require $S = \pm \hat  P_\pi S^\dagger \hat P_\pi$ (either sign) for the $\Sy_2$, $\An_2$ or $(\Fu, \overline{\Fu})$ or  $S = \pm  \hat  P_\pi S^T  \hat P_\pi$ (either sign) for the $\Ad$ or $(\Fu, \Fu)$. Some, but not all, spurions obey these requirements. The spurions used in the next section to generate an example of potential have been chosen to satisfy these invariance requirements.

The second transformation of interest, $G_\pi$, is realized as $ U \to \hat G_\pi U^T  \hat G_\pi^\dagger$ and gives non trivial results only for $\Ccoset$ since in the other two cases $U^T = \pm U$ (see Appendix~B). For the $\Ccoset$ case we choose, following~\cite{Ma:2015gra}

{\footnotesize
\beq
 \hat G_\pi = i \left(\begin{array}{cccc}
 0 & -1 & 0 & 0 \\
 1 & 0 & 0 & 0 \\
 0 & 0 & 0 & 1 \\
 0 & 0 & -1 & 0
\end{array} \right).
\eeq}

\noindent This transformation is interesting because it is also a symmetry of the WZW term and it may be preserved at the quantum level in the UV theory. If so, the lightest neutral pNGBs odd under it (a linear combination of $\phi_0$, $N_0$, $h'$ and $A'$) could be a Dark Matter candidate.

\subsection{Mass spectrum}

Now that we have seen what the main contributions to the potential are and how to compute them, we present a couple of examples of mass spectrum based on a particular choice of spurions. This is not in any way a prediction of the models, it is merely presented to make the previous discussion more concrete and to show qualitatively how a mass spectrum could look like. We consider potentials that depend on three of the dimensionless constants $B_i$, to be specified below. We trade one linear combination for the misalignment angle $\sin\zeta = v/f$, measuring the amount of fine-tuning in the model. A second combination is fixed by imposing the mass of the Higgs boson to be at its measured value \cite{HiggsDiscovery} of 125~GeV. The third combination is left free and varying it gives possible examples for the mass spectrum.

As a first example, consider the $\Rcoset$ model with a potential
\beqs
     V &=& -B_1\; f^4 \; \tr(g^2 T_L^a U T_L^a U^\dagger + g'^2 T_R^3 U T_R^3 U^\dagger) +
     B_2 \; f^4 \; \tr( S_{t_R} U S_{t_R}^* U^*) \nn\\ &&+
     B_3 \; f^4 \; \tr( S_{t_L} U S_{t_L}^* U^* +  S_{b_L} U S_{b_L}^* U^*)
\eeqs
where we have chosen the spurion for $t_R$ to be in the $({\mathbf{1}}, {\mathbf{1}})$ component of the decomposition of the $\Ad$ irrep and the spurion for $(t_L, b_L)$ to be in one of the two $({\mathbf{2}}, {\mathbf{2}})$ components with $T_R^3 = -1/2$ in order for $b_L$ to obey the custodial relations $T_L(T_L+1) = T_R(T_R+1)$ and $T_L^3 = T_R^3$
{\footnotesize \beqs&&
        S_{t_R} =\frac{1}{2 \sqrt 5} \left(
\begin{array}{ccccc}
 1 & 0 & 0 & 0 & 0 \\
 0 & 1 & 0 & 0 & 0 \\
 0 & 0 & 1 & 0 & 0 \\
 0 & 0 & 0 & 1 & 0 \\
 0 & 0 & 0 & 0 & -4
\end{array}
\right), \quad S_{t_L} =\frac{1}{2 \sqrt 2} \left(
\begin{array}{ccccc}
 0 & 0 & 0 & 0 & 0 \\
 0 & 0 & 0 & 0 & 0 \\
 0 & 0 & 0 & 0 & -1-i \\
 0 & 0 & 0 & 0 & 1-i \\
 0 & 0 & 1+i & -1+i & 0
\end{array}
\right), \nn\\ &&\quad S_{b_L} =\frac{1}{2 \sqrt 2} \left(
\begin{array}{ccccc}
 0 & 0 & 0 & 0 & 1-i \\
 0 & 0 & 0 & 0 & -1-i \\
 0 & 0 & 0 & 0 & 0 \\
 0 & 0 & 0 & 0 & 0 \\
 -1+i & 1+i & 0 & 0 & 0
\end{array}
\right).
\eeqs}

\begin{figure}[t]
\begin{center}
\includegraphics[width=0.32\textwidth]{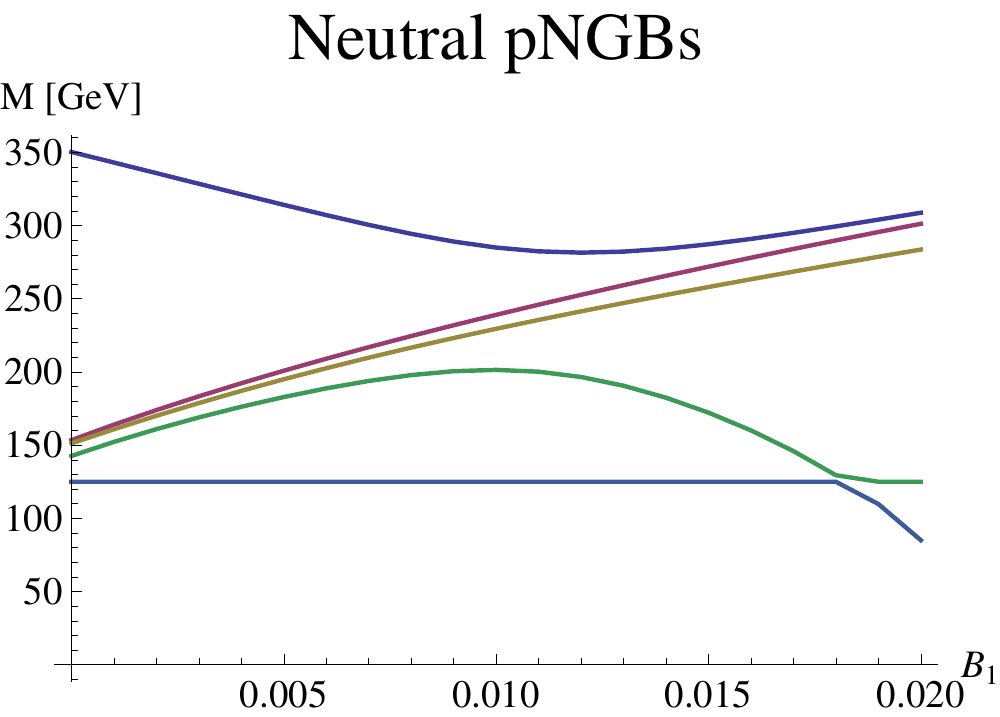} \includegraphics[width=0.32\textwidth]{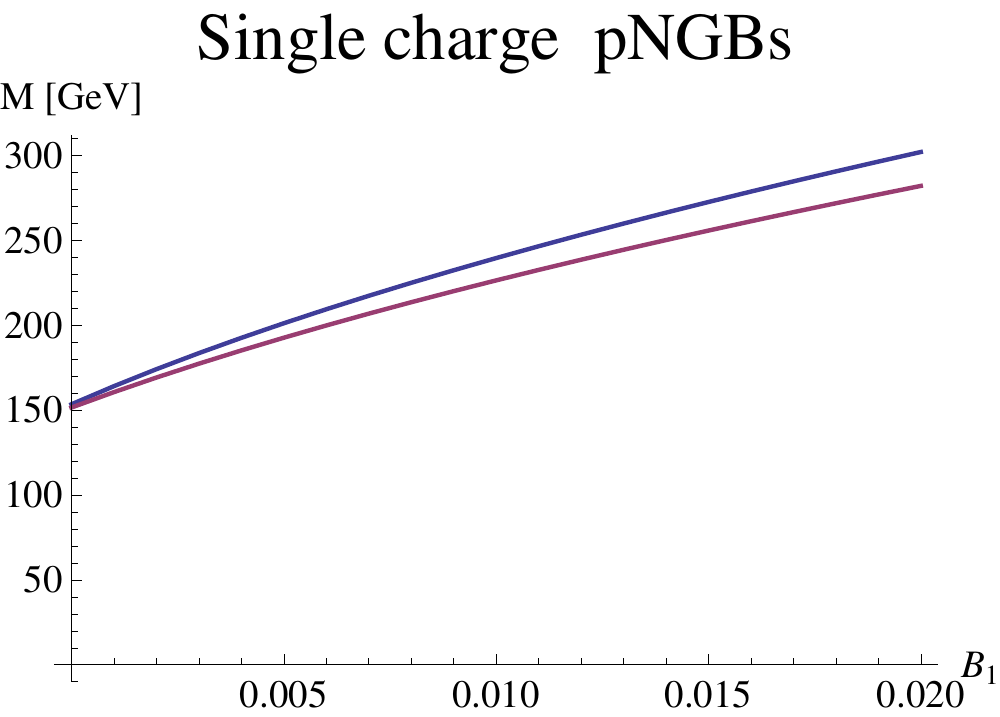}
\includegraphics[width=0.32\textwidth]{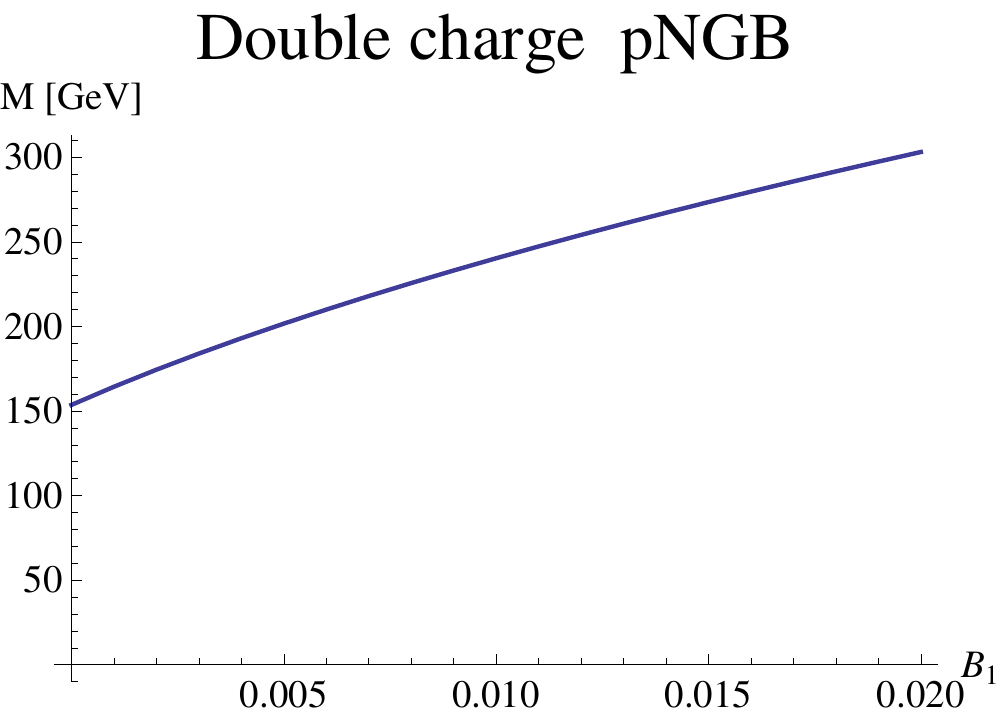}
\caption{Prototypical spectrum for the $\Rcoset$ model with $f = 800$~GeV}
\label{spectrumSOlow}
\end{center}
\end{figure}

\begin{figure}[t]
\begin{center}
\includegraphics[width=0.32\textwidth]{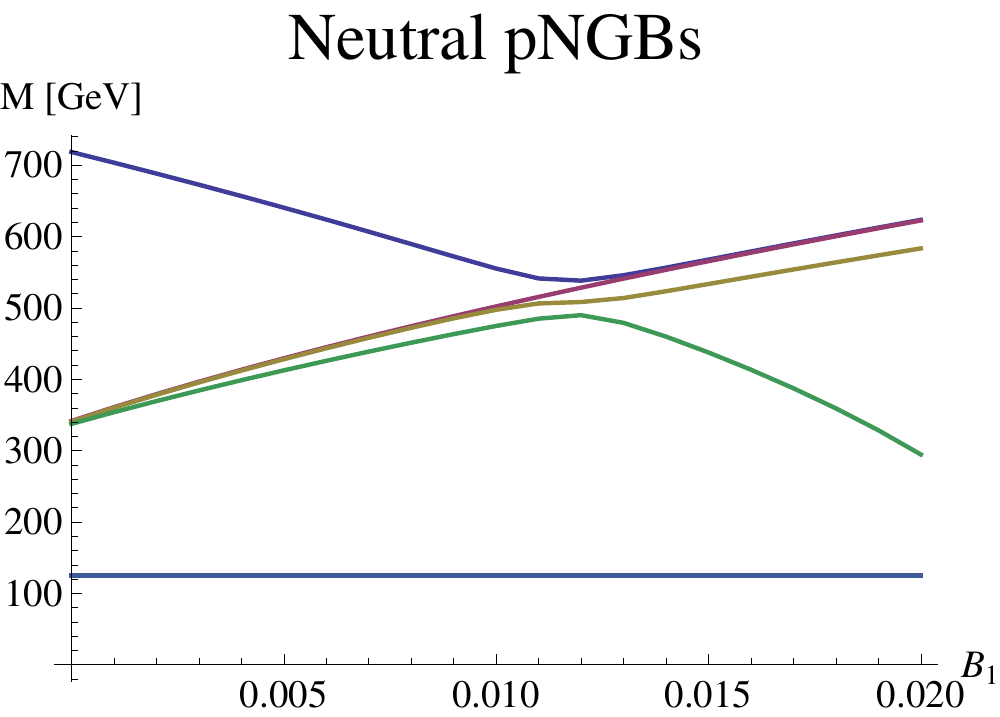} \includegraphics[width=0.32\textwidth]{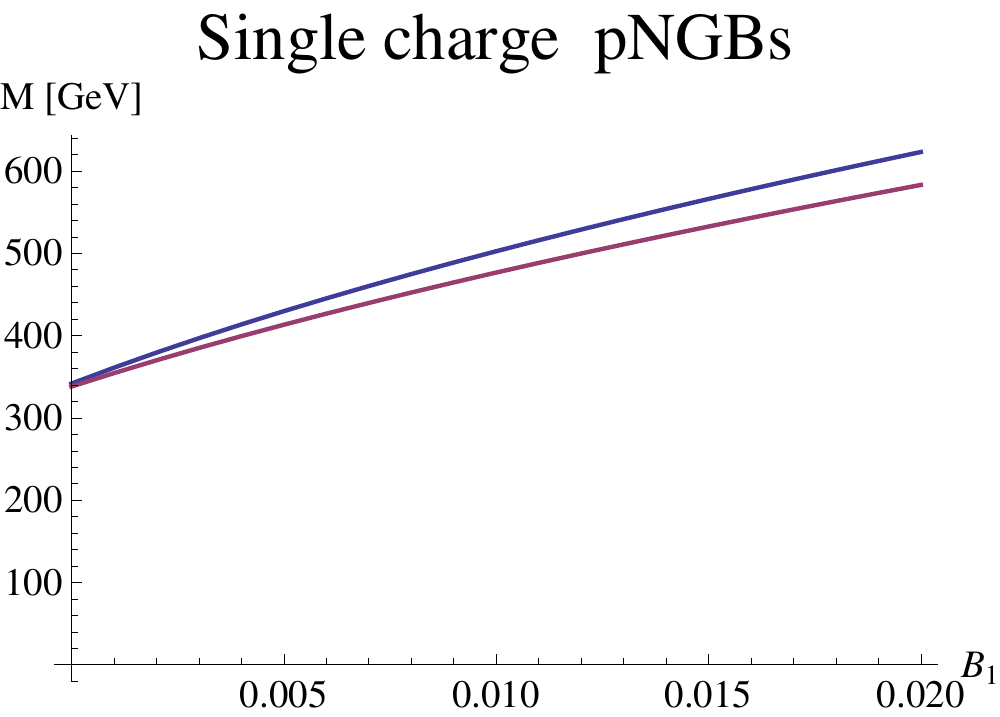}
\includegraphics[width=0.32\textwidth]{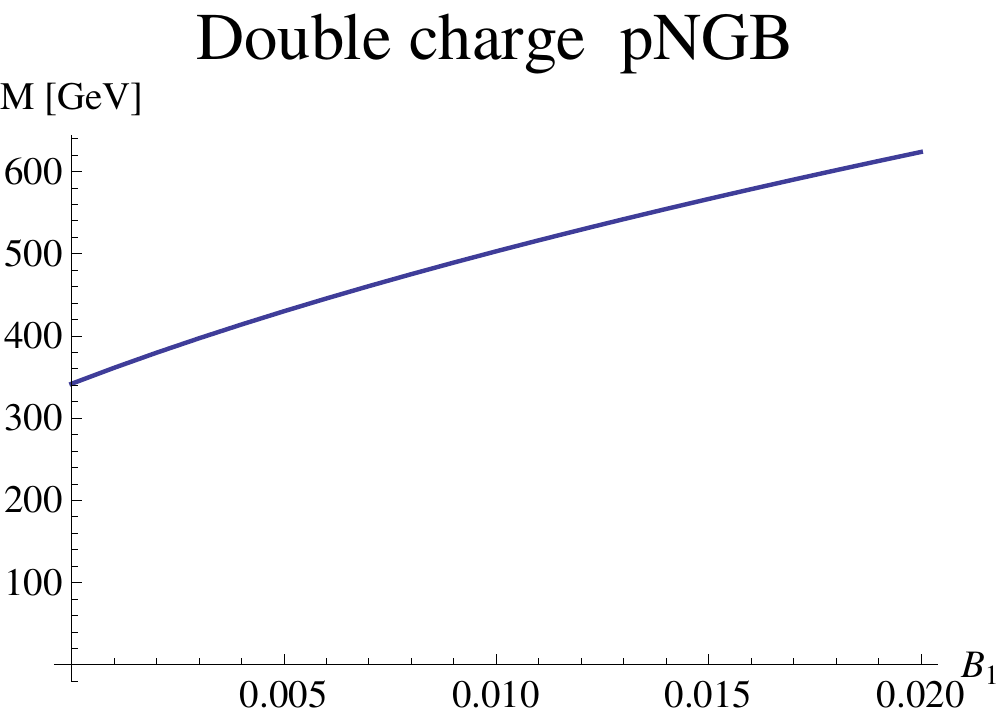}
\caption{Prototypical spectrum for the $\Rcoset$ model with $f = 1600$~GeV}
\label{spectrumSOhigh}
\end{center}
\end{figure}

Setting $f = 800$~GeV and $f = 1600$~GeV, solving the constraints and varying $B_1$ we obtain the spectra in Figure~\ref{spectrumSOlow} and~\ref{spectrumSOhigh} respectively.

Moving on to $\Ccoset$, we chose to present the mass spectrum induced by the following potential, consisting of the contributions from the gauge fields, some bare masses and a LH third family, assumed to give the dominant contribution.

\beqs
     V &=& -B_1\; f^4 \; \tr(g^2 T_L^a U T_L^a U^\dagger + g'^2 T_R^3 U T_R^3 U^\dagger) -
     B_2 \; f^4 \; \tr(  U + U^*) \nn\\ && +
     B_3 \; f^4 \;\left( \tr( S_{t_L}^\dagger U) \tr(S_{t_L} U^\dagger)  +  \tr( S_{b_L}^\dagger U) \tr(S_{b_L} U^\dagger) \right).
\eeqs
The spurions for the LH quarks are chosen to belong to one of the $({\mathbf{2}}, \overline{\mathbf{2}})$ of $SU(2)_L\times SU(2)_R$ found in the decomposition of $({\mathbf{4}}, \overline{\mathbf{4}})$
{\footnotesize \beq
     S_{t_L} = \frac{1}{\sqrt 2}\left(
\begin{array}{cccc}
 0 & 0 & 1 & 0 \\
 0 & 0 & 0 & 0 \\
 0 & 0 & 0 & 0 \\
 0 & -1 & 0 & 0
\end{array}
\right), \quad
     S_{b_L} = \frac{1}{\sqrt 2} \left(
\begin{array}{cccc}
 0 & 0 & 0 & 0 \\
 0 & 0 & -1 & 0 \\
 0 & 0 & 0 & 0 \\
 -1 & 0 & 0 & 0
\end{array}
\right).
\eeq}

\begin{figure}[t]
\begin{center}
\includegraphics[width=0.36\textwidth]{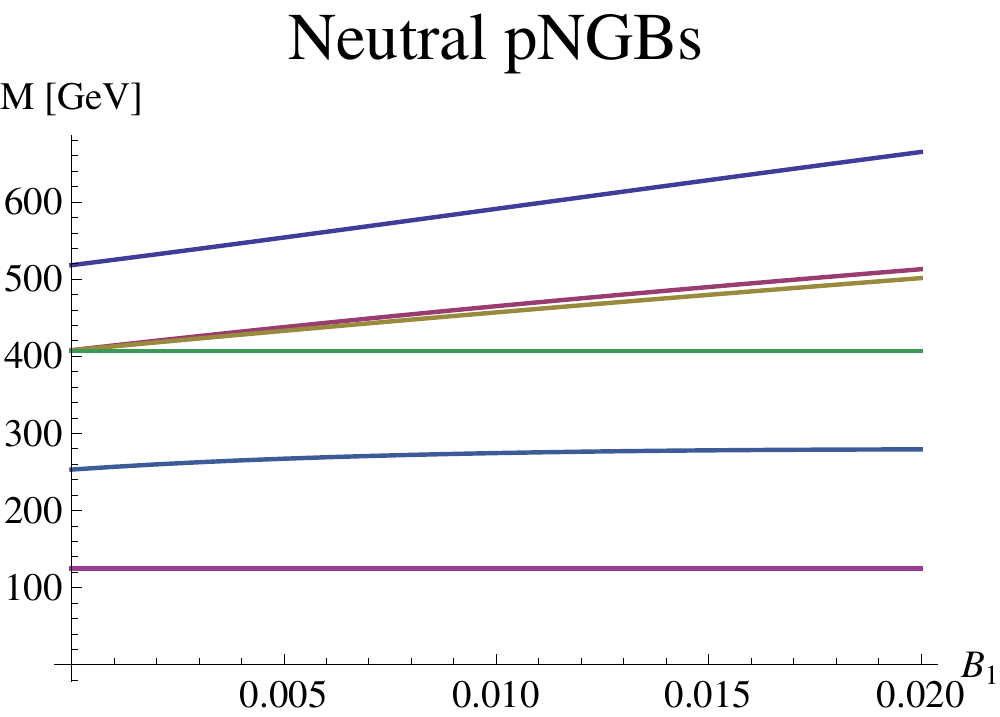} ~~~~~~~~~~~~~~~\includegraphics[width=0.36\textwidth]{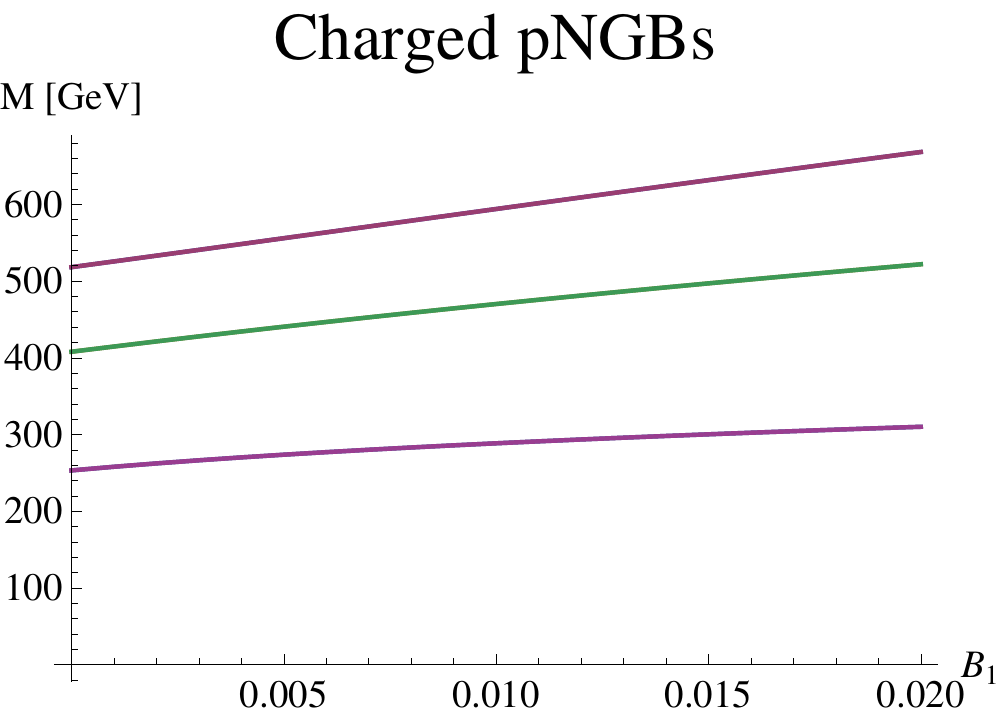}
\caption{Prototypical spectrum for the $\Ccoset$ model with $f = 800$~GeV}
\label{spectrumSUlow}
\end{center}
\end{figure}

\begin{figure}[t]
\begin{center}
\includegraphics[width=0.36\textwidth]{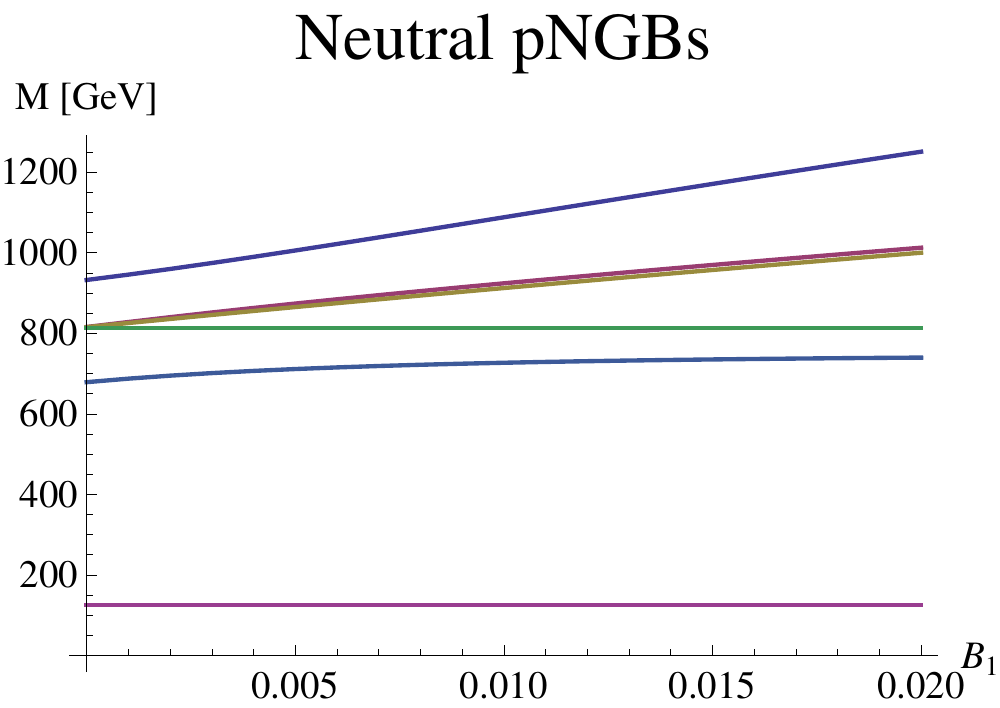}~~~~~~~~~~~~~~~ \includegraphics[width=0.36\textwidth]{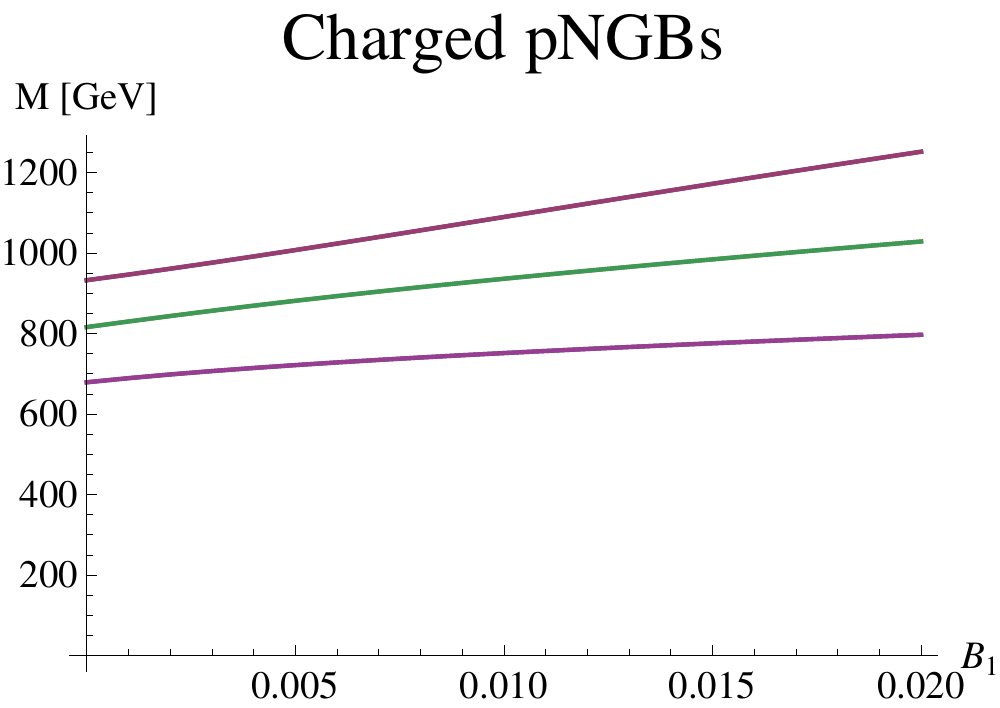}

\caption{Prototypical spectrum for the $\Ccoset$ model with $f = 1600$~GeV}
\label{spectrumSUhigh}
\end{center}
\end{figure}

The representative spectra for $f = 800$~GeV and $f = 1600$~GeV are given in in Figure~\ref{spectrumSUlow} and~\ref{spectrumSUhigh} respectively.

Not much needs to be done for the remaining $\PRcoset$. The $\eta$ is the only pNGB particle other than the Higgs in our current approach its mass is essentially a free parameter. A full discussion of this case is given in~\cite{Gripaios:2009pe}.

\subsection{Couplings involving pNGBs}

The trilinear vertex $\pi \pi' V$ between two generic EW pNGBs and an EW vector boson is encoded in the structure of the currents. With the usual shorthand $\pi_1^*\overleftrightarrow\partial_\mu \pi_2 = \pi_1^*\partial_\mu \pi_2 - \partial_\mu \pi^*_1 \pi_2$ we find, for $\Rcoset$ the coupling to the $Z$-boson\footnote{See again Appendix B for notation on the field content of the theory. We set $\sz = \sin\zeta$, $\sw = \sin\theta_W$ etc.}
\beq
    {\mathcal{L}} \supset   \frac{i e}{2\sw \cw}Z^\mu \left( (\cww - \cz)\phi_-^0\overleftrightarrow\partial_\mu\phi_+^0 +
     (\cww+ \cz)\phi_0^-\overleftrightarrow\partial_\mu\phi_0^+ + 2 \cz \phi_+^- \overleftrightarrow\partial_\mu \phi_-^+ +
     2 \cww \phi_-^- \overleftrightarrow\partial_\mu\phi_+^+\right)
\eeq
and that to the $W^\pm$
\beqs
    {\mathcal{L}}\supset   \frac{e}{2\sw}W^{-\mu}&&\!\!\!\!\!\!\!\!\Big( (1+\cz) (\phi_-^+ \overleftrightarrow\partial_\mu  \phi_+^0 +
    \phi_0^0 \overleftrightarrow\partial_\mu  \phi_0^+ - \phi_-^0 \overleftrightarrow\partial_\mu  \phi_+^+ )\nn\\
    &&\!\!\!\!\!\!\!\!\! -(1-\cz) (\phi_0^0 \overleftrightarrow\partial_\mu  \phi_+^0  +  \phi_+^- \overleftrightarrow\partial_\mu  \phi_0^+ -
     \phi_0^- \overleftrightarrow\partial_\mu  \phi_+^+) \Big) + \mbox{ h.c.}
\eeqs

For $\Ccoset$ we find instead, in agreement with the results of~\cite{Ma:2015gra}
\beq
    {\mathcal{L}} \supset   \frac{i e}{2\sw \cw}Z^\mu \left( (\cww - \cz)N_- \overleftrightarrow\partial_\mu N_+ +
    (\cww + \cz)\phi_- \overleftrightarrow\partial_\mu \phi_+ + \cww H'_- \overleftrightarrow\partial_\mu H'_+ +
     i \cz A' \overleftrightarrow\partial_\mu h' \right)
\eeq
for the $Z$ couplings, and
\beq
    {\mathcal{L}}\supset  - \frac{i e}{2\sw}W^{-\mu}\Big( (1-\cz) N_0 \overleftrightarrow\partial_\mu N_+
    + (1 + \cz) \phi_0 \overleftrightarrow\partial_\mu \phi_+ - \cz h' \overleftrightarrow\partial_\mu H'_+
    + i A' \overleftrightarrow\partial_\mu H'_+  \Big) + \mbox{ h.c.}
\eeq
for the $W^\pm$ couplings. The electromagnetic coupling is of course always given by $i e q_\pi A^\mu\, \pi^* \overleftrightarrow\partial_\mu \pi $ for any of the pNGBs $\pi$ of charge $q_\pi$.

In all three cases the Higgs boson $h$ does not mix with the other pNGBs and its couplings to the vector bosons at tree level are~\footnote{For uniformity we have chosen to normalize all three cases according to $m_W = \frac{1}{2}g f\sin\zeta$, implying $v = f\sin\zeta = 246$~GeV. This is different from the normalization of $f$ used in~\cite{Ferretti:2014qta}. Our $h$ is already shifted to have zero v.e.v. and is canonically normalized.}:
\beq
    {\mathcal{L}} = \frac{1}{4} g^2 f \szz h W^{+\mu} W_\mu^- +
     \frac{1}{8} (g^2 + g'^2) f \szz h Z^\mu Z_\mu +
      \frac{1}{4} g^2  \czz h^2 W^{+\mu} W_\mu^- +
     \frac{1}{8} (g^2 + g'^2)  \czz h^2 Z^\mu Z_\mu.
\eeq

The model $\PRcoset$ only contains the $\eta$ as an additional pNGB. Its trilinear couplings vanish and at quartic level it can easily be written down:
\beq
    {\mathcal{L}} = - \frac{1}{4} g^2  \sz^2 \eta^2 W^{+\mu} W_\mu^- -
     \frac{1}{8} (g^2 + g'^2)\sz^2 \eta^2 Z^\mu Z_\mu.
\eeq

For the quartic couplings in the remaining models we refer to Appendix~C.

The $P_\pi$-parity odd pNGBs can decay to the transverse part of the vector bosons via the anomaly term yielding a vertex $\pi V V'$. This can be extracted from the WZW term~\cite{WZW} by considering the piece containing one pNGB and two vector bosons. The relevant term is given in~\cite{Kaymakcalan:1983qq} in the elegant language of differential forms
\beqs
    S_{\mathrm WZW} \supset \frac{i \dim(\psi)}{48\pi^2}\int\tr\!\!\!&\bigg(&\!\!\!\!dA_L A_L dU U^\dagger + A_L dA_L dU U^\dagger
    + dA_R A_R U^\dagger dU + A_R dA_R U^\dagger dU \nn\\ &&\!\!\!\!- dA_L dU A_R U^\dagger + dA_R dU^\dagger A_L U \bigg).
\eeqs

For $\PRcoset$ we set $A_L = A$, $A_R = -A^T = -\epsilon_0 A \epsilon_0$ and $U = \Omega \exp(2 \sqrt{2} i \Pi/f)\epsilon_0 \Omega^T$. Expanding to first order in the pNGBs and integrating by parts yields
\beq
      S_{\mathrm WZW} \supset  \frac{ \dim(\psi)}{16\pi^2 f}\cz \int\eta\bigg(\frac{g^2 - g'^2}{2} Z_{\mu\nu}\tilde Z^{\mu\nu} +
       g g'  F_{\mu\nu}\tilde Z^{\mu\nu} + g^2  W^+_{\mu\nu}\tilde W^{-\mu\nu}\bigg){\mathrm{d}}^4x. \label{WZWeta}
\eeq

For $\Ccoset$ we set $A_L = A_R = A$ and $U = \Omega \exp(2 \sqrt{2} i \Pi/f) \Omega$. Expanding to first order in the pNGBs and integrating by parts we find exactly the same expression as (\ref{WZWeta}). This was found in~\cite{Ma:2015gra} and it is due to the extra symmetry $G_\pi$, defined in Section~3.2, present in this case. In particular, no terms involving the pNGB $\phi$ and $N$ arise in this model.

On the contrary, for the coset $\Rcoset$, we need to set $A_L = A$, $A_R = -A^T = A$ and $U = \Omega \exp(2 i \Pi/f) \Omega^T$. Here, no additional symmetry is present and all the pNGBs other than the Higgs boson appear in the WZW action. In this case, the trilinear anomalous couplings are presented in Appendix~C.

\begin{figure}[t]
\begin{center}
\includegraphics[width=\textwidth]{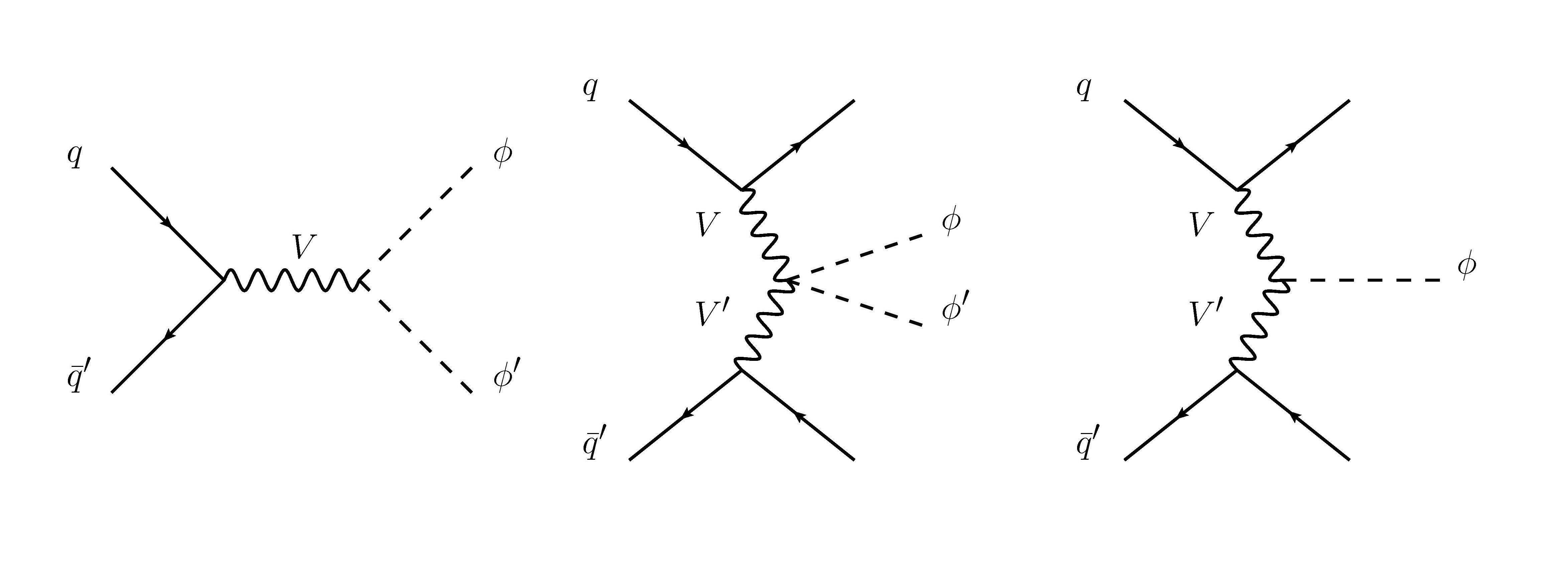}
\caption{Production modes for the EW pNGBs. From left to right: Drell-Yan (DY), Vector Boson Fusion pair production via a renormalizable quartic interaction (VBFr) and Vector Boson Fusion via the anomaly (VBFa).}
\label{productionmodes}
\end{center}
\end{figure}

There are three possible production modes to be considered for these EW pNGBs, see Figure~\ref{productionmodes}. Two of them are pair production modes, one by an off-shell vector boson in the $s$-channel -- Drell-Yan production (DY) -- and the other by vector boson fusion via a renormalizable four boson interaction (VBFr). The third one is a single production mode by vector boson fusion via the anomaly (VBFa).

Perhaps surprisingly, VBFr tends to give a larger contribution than DY. Consider the interesting case of the doubly charged pNGB $\phi_+^+$ present in $\Rcoset$. (A model in which such a particle is present as an elementary object is the Georgi-Machacek model~\cite{Georgi:1985nv}.)
The tree level production can be easily estimated with {\tt MadGraph} and {\tt FeynRules}~\cite{software} yielding, at 13~TeV for a mass of 500~GeV and $f=800\mbox{ GeV}$: $\sigma_{\mathrm{DY}}(\phi_+^+ \phi_-^-) = 1.3~\mathrm{fb}$ and $\sigma_{\mathrm{VBFr}}(\phi_+^+ \phi_-^-) = 3.0~\mathrm{fb}$. 

The single production of the doubly charged pNGBs  via VBFa is totally negligible in this case: $\sigma_{\mathrm{VBFa}}(\phi_+^+) = 2.3\times 10^{-5}~\mathrm{fb}$. This last statement is no longer true for other pNGBs. For instance, in the case of the $\eta$ of $\PRcoset$, (and a particle with exactly the same couplings is present in $\Ccoset$ as well), with the same parameters as before, the double production is now negligible: $\sigma_{\mathrm{DY}}(\eta\eta) = 0$ (impossible) and $\sigma_{\mathrm{VBFr}}(\eta\eta) = 2.0\times 10^{-2}~\mathrm{fb}$, while $\sigma_{\mathrm{VBFa}}(\eta)$ is of the order of a few fb depending on the specific value of the anomaly.

The reason for this different behavior is due to the fact that the VBF diagrams that contribute the most are those where a photon is allowed to be present.
For this same reason, the single charge pNGBs have non negligible cross section for all processes and the single production mode becomes relevant at higher masses. We have not tried to pin down the exact range of masses where one production mode is expected to be dominant with respect to the others because this depends on the details of the models such as mixing, which is not an issue for the $\eta$ of $\PRcoset$ or the $\phi_+^+$ of $\Rcoset$. However, given that
$\sigma_{\mathrm{VBFr}}(\phi_+^+ \phi_-^-)$ and $\sigma_{\mathrm{VBFa}}(\eta)$ are roughly comparable for masses of 500~GeV, we expect the cross-over region to be within the energy range of the LHC.

\begin{figure}[t]
\begin{center}
\includegraphics[width=0.25\textwidth]{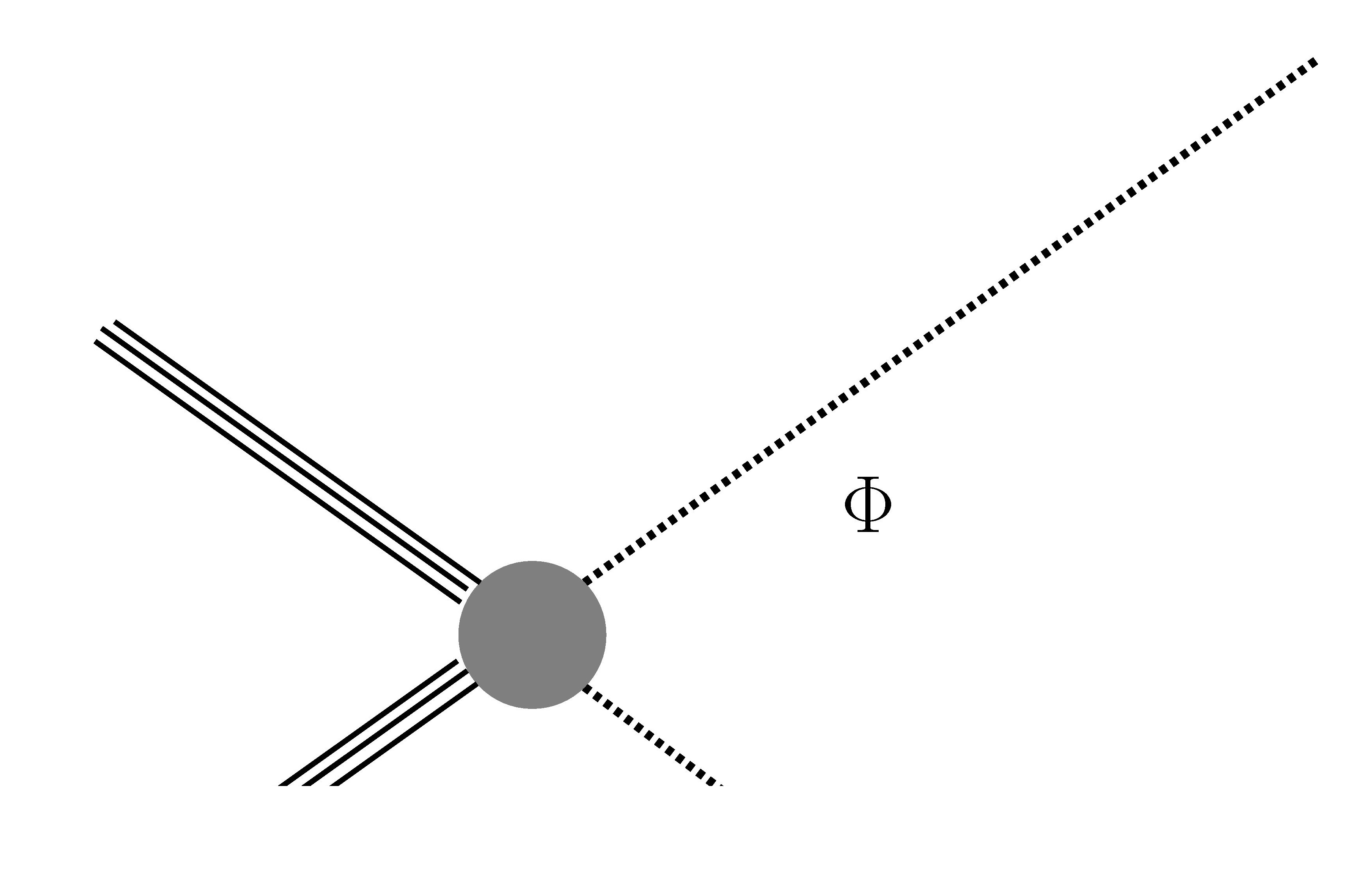}~~~~
\includegraphics[width=0.25\textwidth]{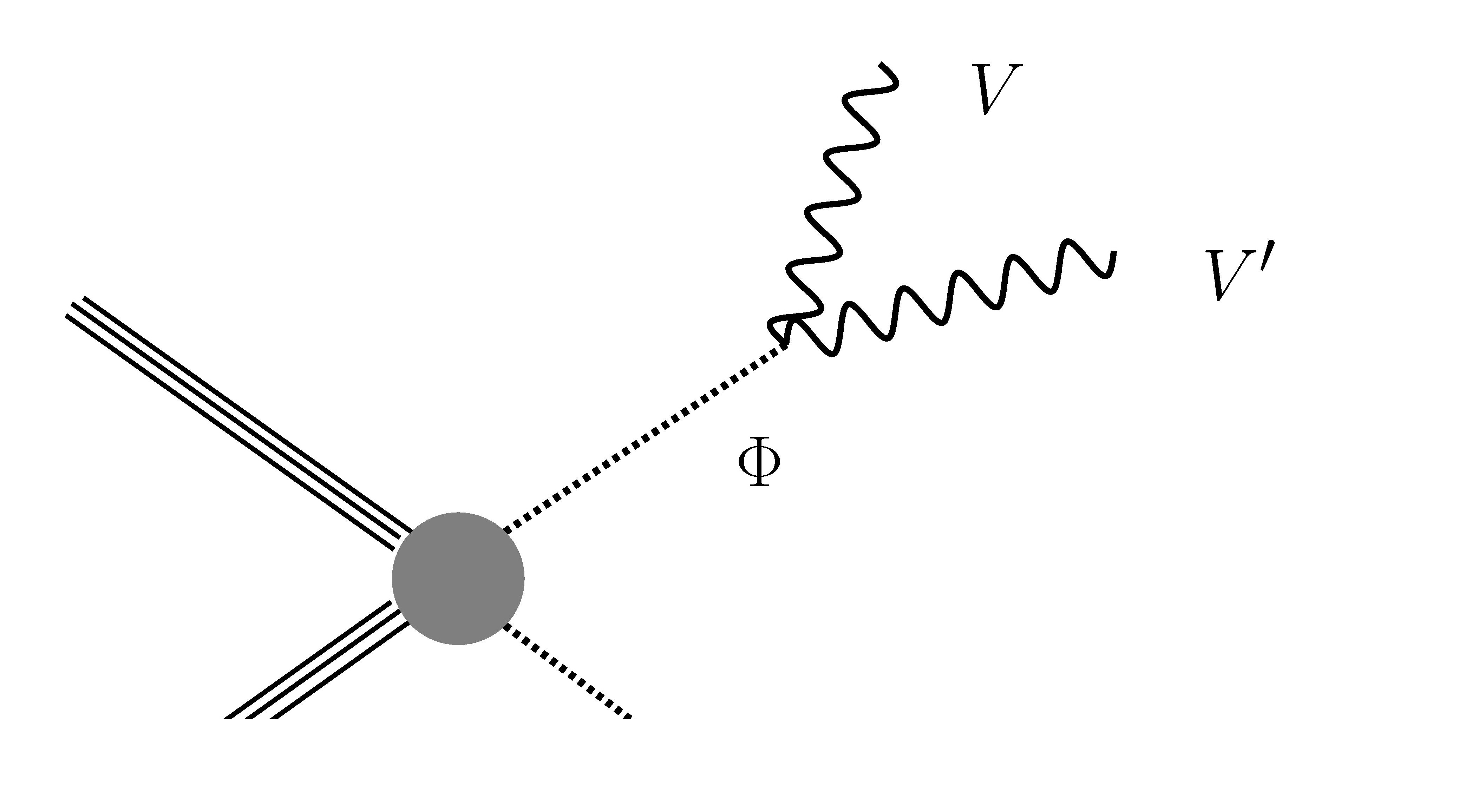}~~~~
\includegraphics[width=0.25\textwidth]{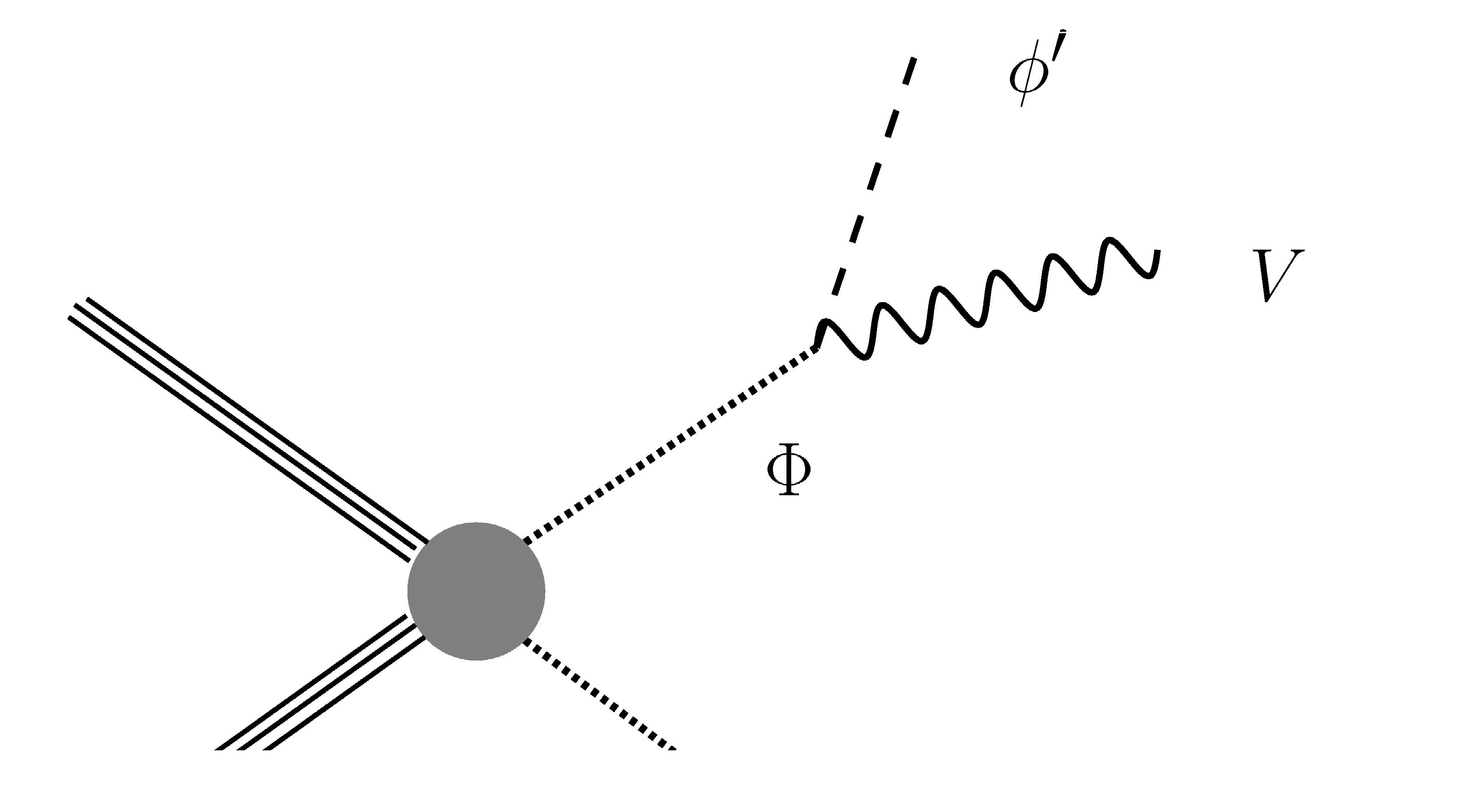}~~~~
\includegraphics[width=0.25\textwidth]{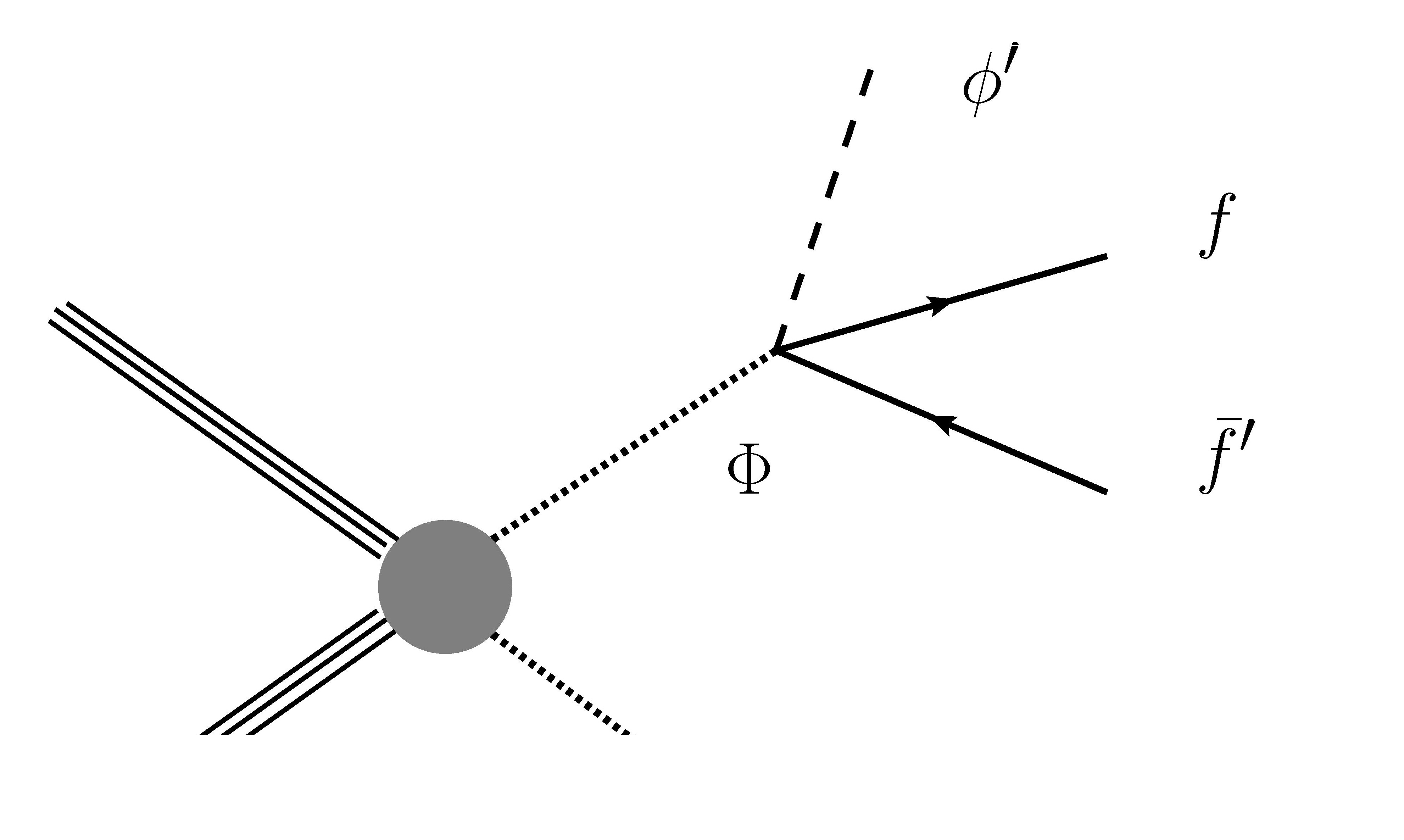}
\caption{After production, an EW pNGB $\Phi$ can be collider stable, decay to two vector bosons $V, V'$ via the anomaly, decay to a lighter pNGB $\phi'$ plus a vector boson $V$ or to a pair of fermions $f, \bar f'$ plus a lighter pNGB $\phi'$.}
\label{decayEW}
\end{center}
\end{figure}

The phenomenology of the cosets $\Ccoset$ and  $\Rcoset$ is potentially very rich (some would say too rich...). Once produced, the EW pNGBs chain decay to lighter ones plus a SM vector boson, if kinematically allowed, or a pair of SM fermions.
In the $\Rcoset$ case, the lightest EW pNGB decays to two SM vector bosons via the anomaly. (This may actually become the dominant decay mode for heavier pNGBs as well if the spectrum is squeezed, $\Delta m\lesssim 10 {\mbox{ GeV}}$.)
In the $\Ccoset$ case, the lightest pNGB odd under $G_\pi$ is collider stable under our assumptions and thus leads to missing energy or charged heavy tracks depending on its charge. If its decay into SM fermions is totally forbidden, it could even be a dark matter candidate~\cite{Ma:2015gra}. This is in the spirit of~\cite{Frigerio:2012uc} although their candidate for dark matter (the $\eta$ of $\PRcoset$) is not viable for our UV completions because it decays through the anomalous couplings.
(For pNGB dark matter see also~\cite{Kim:2016jbz}. Additional dark matter candidates have been conjectured to arise from the topological structure of similar cosets~\cite{Joseph:2009bq}.) A pictorial description of the various possibilities is given in Figure~\ref{decayEW}.

\section{Top partners and colored mesons}

We now turn to the discussion of objects carrying color, that is, bound states containing some of the constituents $\chi$.

As we mentioned in the introduction, top-partners are realized via fermionic tri-linears in the hyperquarks. These can be of type $\psi\chi\psi$ or $\chi\psi\chi$ depending on the type of model under consideration, as shown in Appendix~A. So far we have been somewhat sloppy in indicating the structure of these objects, now it is time to be more specific.

We need at least six new fermions ``$\chi$'' in order to embed the color group into the associated global symmetry group in an anomaly-free way.
In the case of a complex irrep, leading to $\Ccolor$, the $\chi$s are already naturally divided into 3 $\chi$ and 3 $\tilde\chi$ transforming in conjugate irreps $R_\chi$ and $\overline{R}_\chi$ of $\GHC$ as well as the ${\mathbf{3}}$ and $\overline{{\mathbf{3}}}$ of color $SU(3)_c$. (As always, we are using Weyl LH notation.)

Even in the other two cases (real or pseudo-real irreps), it is still convenient to split the 6 fermions into a ${\mathbf{3}}+ \overline{{\mathbf{3}}}$ of $SU(3)_c$. In these cases we allow ourselves the following notational ambiguity
\beq
        \chi\in{\mathbf{6}} \equiv \left(
                                     \begin{array}{c}
                                       \chi \\
                                       \tilde\chi \\
                                     \end{array}
                                   \right)\in
                                     \left(
                                     \begin{array}{c}
                                       {\mathbf{3}}_X \\
                                       {\overline{\mathbf{3}}}_{-X} \\
                                     \end{array}
                                   \right)  \label{abuse}
\eeq
for ease of notation. Note that these fermions must carry not only the color quantum numbers but also the additional $U(1)_X$ charge needed to obtain the proper weak hypercharge $Y = X + T_R^3$ for the top partners. The allowed values of $X$ can be found looking at the construction of the top-partners as follows.

Consider the case where the top-partners are of type $\chi\psi\chi$. Using the notation (\ref{abuse}), we can generally construct at most three types of LH objects transforming in the ${\mathbf{3}}$. They are contained in the products $\tilde\chi\psi\tilde\chi$, $\tilde\chi\psi^\dagger\chi^\dagger$, $\chi^\dagger\psi\chi^\dagger$, where we used the fact that $\overline{\mathbf{3}}\times \overline{\mathbf{3}} = \overline{\mathbf{6}} + \mathbf{3}$.
Identifying the $T_L=T_R=0$ component with the partner of $t_R$ we see that we must chose $X(\chi) = -1/3$ and $B(\chi)=-1/6$ (baryon number) for the constituents $\chi$.

Now, still within the $\chi\psi\chi$ case, if the $\GHC$ irrep for the $\chi$ in question is real, giving rise to the coset $\Rcolor$, this leads to colored pNGBs $\chi\chi \in {\mathbf{6}}_{-2/3}$ of baryon number $1/3$, as well as $\tilde\chi\tilde\chi \in \overline{{\mathbf{6}}}_{+2/3}$ of baryon number $-1/3$ and the ever-present $\tilde\chi\chi \in {\mathbf{8}}_0$ of baryon number $0$.
If  the $\GHC$ irrep is pseudo-real, giving rise to the coset $\PRcolor$, then the pNGB mesons are $\chi\chi\in \overline{{\mathbf{3}}}_{-2/3}$ etc. with the same baryon number assignments as before.

If instead the top partners are of type $\psi\chi\psi$, then the $\chi$ and $\tilde\chi$ in (\ref{abuse}) must be in
the ${\mathbf{3}}_{+2/3}+ \overline{{\mathbf{3}}}_{-2/3}$ of $SU(3)_c \times U(1)_Y$ with baryon number $\pm 1/3$, leading, for a real irrep, to mesons
$\chi\chi \in {\mathbf{6}}_{4/3}$ of baryon number $2/3$ and its complex conjugate plus the usual $\tilde\chi\chi \in {\mathbf{8}}_0$. From Appendix~A we see that no pseudo-real cases exist when the top-partners are of type $\psi\chi\psi$. The case in which the $\chi$ are in a complex irrep only leads to the neutral meson $\tilde\chi\chi \in {\mathbf{8}}_0$ without baryon number.

The masses for these colored objects should be in the multi TeV range getting contributions from gluon loops and possibly bare masses for $\chi$ but they could still be in the discovery range of LHC. The octets decay mostly to two gluons via the anomaly term
\beq
     {\mathcal{L}}_{WZW} \supset \frac{g_s^2 \dim{\chi}}{16\pi^2 f_c} d^{ABC}\Pi^A G^B_{\mu\nu}\tilde G^{C\mu\nu}  \label{anoglue}
\eeq
but there is no such term available for the triplet or the sextet. Preserving $P_\pi$-parity, we can let them cascade to the lighter EW pNGBs via interactions of type $\pi q q' \phi$ where $q$ and $q'$ are SM quarks and $\phi$ is an appropriate EW pNGB with the right quantum numbers.
If we allow for interactions violating $P_\pi$-parity, we do not need this additional pNGB. Summarizing, we have therefore the following three possibilities, in addition to the octet:
\bit
    \item{Case a)} $\chi$ in a real irrep and top-partners of type $\chi\psi\chi$. This gives rise to mesons $\pi$ in the ${\mathbf{6}}_{-2/3}$ of $SU(3)_c\times U(1)_Y$ of baryon number $-1/3$. They can decay via $\Delta B=1$ couplings
\beq
     \pi^{*ab} Q_{La} Q_{Lb} \phi,\quad  \pi^{*ab} u_{Ra} u_{Rb} \phi ,\quad  \pi^{*ab} d_{Ra} u_{Rb} \phi ,\quad  \pi^{*ab} d_{Ra} d_{Rb}\phi
\eeq
where we denoted explicitly only the color index. The various EW pNGBs $\phi$ appearing in the vertex must be such that the particular vertex is invariant under the full SM gauge group. In the case of $Q_L Q_L$ coupling, we have the option of coupling to a $SU(2)_L$ triplet or a singlet, making the quark flavor indices symmetric or anti-symmetric respectively. In all gory details for the triplet:
$\pi^*_{ab} Q_{La}^{\alpha f i} Q_{Lb\alpha}^{f' j} \phi_{ij}$, symmetric in the exchange of $f f'$.
In the absence of $P_\pi$-parity we could also consider the term $\pi^{*ab} d_{Ra} d_{Rb}$, symmetric in the flavor indices.

\item{Case b)} $\chi$ in a real irrep and top-partners of type $\psi\chi\psi$. This gives rise to mesons $\pi$ in the ${\mathbf{6}}_{4/3}$ of $SU(3)_c\times U(1)_Y$ of baryon number $2/3$. They can decay via same couplings as case a) but now these couplings are baryon number preserving. Without $P_\pi$-parity one can only make the vertex $\pi^{*ab} u_{Ra} u_{Rb}$, symmetric in flavor.

\item{Case c)} $\chi$ in a pseudo-real irrep and top-partners of type $\chi\psi\chi$.
The extra mesons are now in the ${\mathbf{3}}_{2/3}$ of baryon number $1/3$ and decay via the $\Delta B=1$ interactions
\beq
     \epsilon^{abc}\pi_a Q_{Lb} Q_{Lc} \phi,\quad  \epsilon^{abc}\pi_a u_{Rb} u_{Rc} \phi ,\quad  \epsilon^{abc}\pi_a d_{Rb} u_{Rc} \phi ,\quad  \epsilon^{abc}\pi_a d_{Rb} d_{Rc}\phi
\eeq
with the appropriate EW pNGB. Without $P_\pi$-parity one can construct $\epsilon^{abc}\pi_a d_{Rb} d_{Rc}$ asymmetric in the flavor indices.
\eit

\begin{figure}[t]
\begin{center}
\includegraphics[width=0.4\textwidth]{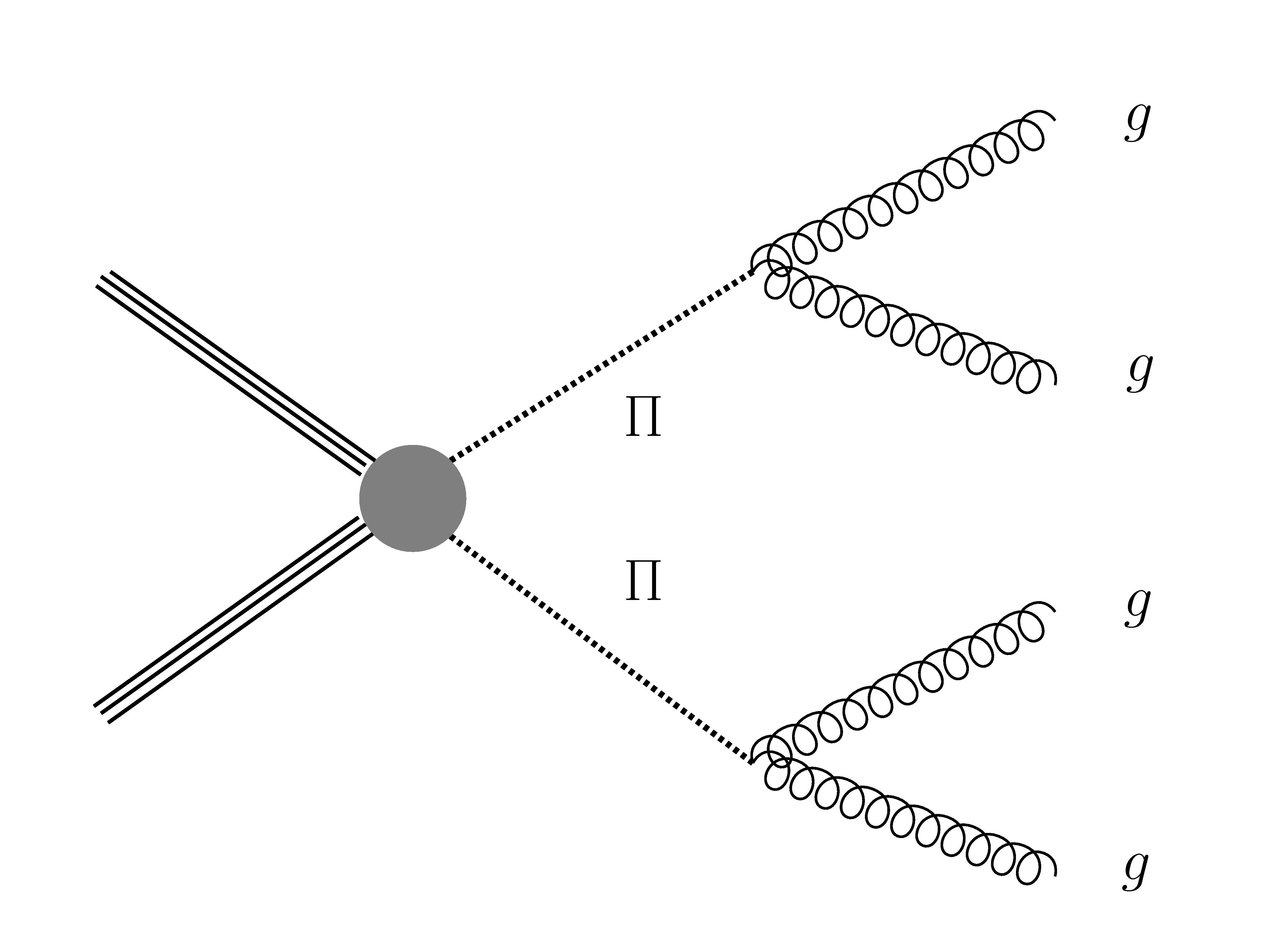}~~~~
\includegraphics[width=0.4\textwidth]{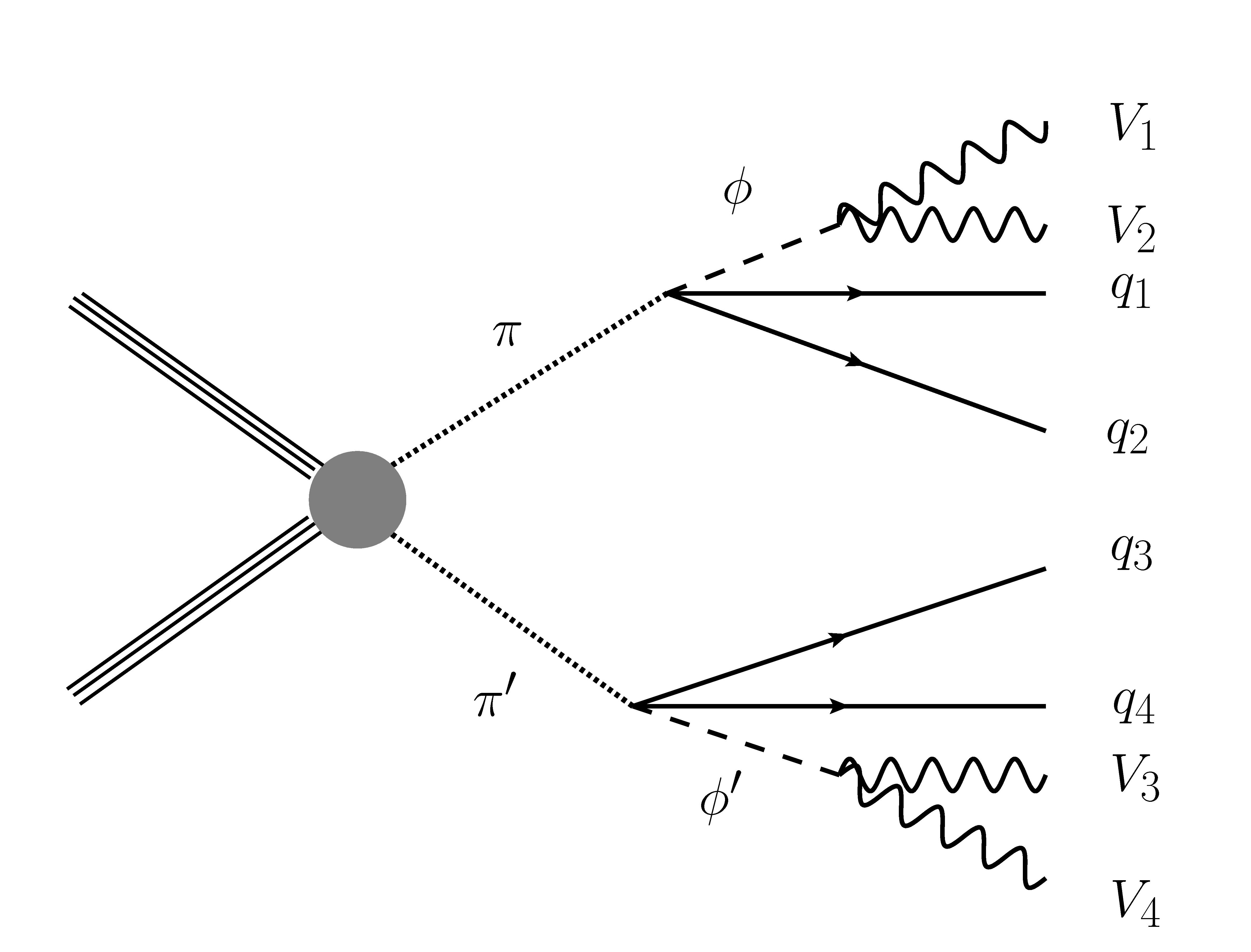}
\caption{Pair production and decay of color octets (left) and triplet/sextet (right). The octet can also be singly produced by the anomalous coupling via gluon fusion.}
\label{coloredplusdecay}
\end{center}
\end{figure}

For all EW cosets there are some pNGBs that can be used to construct some of the couplings, so all the colored sextets and triplets can decay into two jets and an EW pNGB. Note that proton stability is assured since we preserve lepton number. However, the presence of $\Delta B=1$ interactions raises the interesting possibility of neutron-anti-neutron oscillations. (See~\cite{Calibbi:2016ukt} for a recent discussion in the context of RPV-SUSY. Similar scalars objects have been discussed in e.g.~\cite{oai:arXiv.org:0910.1789, Cacciapaglia:2015eqa}). The situation is summarized in Figure~\ref{coloredplusdecay}.

As far as fermionic colored objects go these models predict a slew of additional resonances but all of them, with the possible exception of the top partners, should be out of reach at LHC.

Exotic fermions of higher electric charge also need be taken into consideration. For the almost ubiquitous charge $5/3$ state $X$, the main decay mode targeted by experiments so far is $X \to W \; t$ \cite{Chatrchyan:2013wfa}, but the existence of possible additional charged pNGBs opens alternative channels such as $X \to t \; \phi_0^+$. The presence of doubly charged pNGBs in some constructions might even allow for $X \to b \; \phi_+^+$.

The operator creating the fermionic resonance should acquire a large negative anomalous dimension in the running from $\Lambda_{\mathrm{UV}}$ to $\Lambda$. This has been investigated at the perturbative level in \cite{DeGrand:2015yna} for the class of models in~\cite{Ferretti:2014qta}. More recently \cite{Pica:2016rmv} summarized the results for the QCD case, also within perturbation theory.

\section{Two more pNGBs/ALPs}

A universal feature of all of these models, simply due to the fact that they are constructed out of two different types of fermions, is the existence of two additional neutral pNGBs associated to the abelian axial currents from the axial $U(1)_\psi$ and $U(1)_\chi$. One linear combination of these currents can be taken to be free of $\GHC$ anomalies. The associated pNGB, to be denoted by $a$, will be naturally light and, in absence of further interactions would essentially be a composite axion~\cite{Kim:1984pt} coupling to both gluons and EW gauge bosons via the anomaly
\beq
      {\mathcal{L}} =  \frac{g_s^2 N_s}{16\pi^2 f_a} a G^A_{\mu\nu}\tilde G^{A\mu\nu}
      + \frac{{g'}^2 N_B}{16\pi^2 f_a} a B_{\mu\nu}\tilde B^{\mu\nu} +
      \frac{{g}^2 N_W}{16\pi^2 f_a} a W^i_{\mu\nu}\tilde W^{i\mu\nu}.
\eeq
Since the associated decay constant $f_a$ is much smaller than the possible window of values allowed by the ``invisible-axion'' solution, we must give this particle a mass to avoid the usual constraints.
As in technicolor models~\cite{Farhi:1980xs}, a mass can be obtained from e.g. the four-fermi terms arising at the $\Lambda_{\mathrm{UV}}$ scale of the type ($c_i = {\mathcal{O}}(1)$)
\beq
    H' = -{\mathcal{L}}_{\mathrm{4f}} = \frac{1}{\Lambda^2_{\mathrm{UV}} }\left( c_1 \chi^2 \tilde\chi^2 + c_2 \psi^4 + \dots\right).
\eeq
For typical values of the parameters, using Dashen's formula~\cite{Dashen:1969eg} we estimate
\beq
       m^2_a = \frac{1}{f^2}\langle [Q, [Q, H']]  \rangle \approx \frac{\Lambda^6}{f^2{ \Lambda^2_{\mathrm{UV}}}} \approx \frac{(5\times 10^3~\mathrm{GeV})^6}{(800~\mathrm{GeV})^2 (10^8~\mathrm{GeV})^2} \approx (1.6~\mathrm{GeV})^2
\eeq
but a fairly large range of masses is possible. For instance, Naive Dimensional Analysis would lead to a lower estimate 
$m^2_a \approx \Lambda^2 f^2 /\Lambda^2_{\mathrm{UV}}\approx (40.~\mathrm{MeV})^2$.
This value needs to be raised at least by roughly a factor $\approx 3$ in order not to conflict with the bounds on the visible axion, coming from beam dump experiments (discussed in~\cite{Dobrich:2015jyk}) or $K\to\pi a$ searches~\cite{Anisimovsky:2004hr}. (See also~\cite{Cadamuro:2011fd} for cosmological bounds for ALPS at much higher scale $f$.) This however is easily achieved. In fact, in~\cite{Belyaev:2015hgo} the exciting possibility has been raised that this object is responsible for the 750~GeV bump in the di-photon signal recently reported by ATLAS and CMS~\cite{diphotonATLASCMS}. Such a large mass could be obtained by e.g. adding bare masses for the colored hyperquarks.

The remaining linear combination, to be denoted by $\eta'$, corresponds to the $\GHC$ anomalous current and its associated ``would-be'' Goldstone boson acquires a mass via the 't~Hooft mechanism~\cite{'tHooft:1976up}. The $\eta'$ mass is given the Veneziano-Witten formula~\cite{Veneziano:1979ec, Witten:1979vv} ($N\approx 10$, $\Xi$ the topological susceptibility)
\beq
     m_{\eta'}^2 = \frac{2 N}{f^2} \Xi.
\eeq
that can be naively estimated to be of the same order of a typical resonance. However, subtleties may arise that lower the mass of this object and also make it within reach of the LHC.

\begin{figure}[t]
\begin{center}
\includegraphics[width=0.5\textwidth]{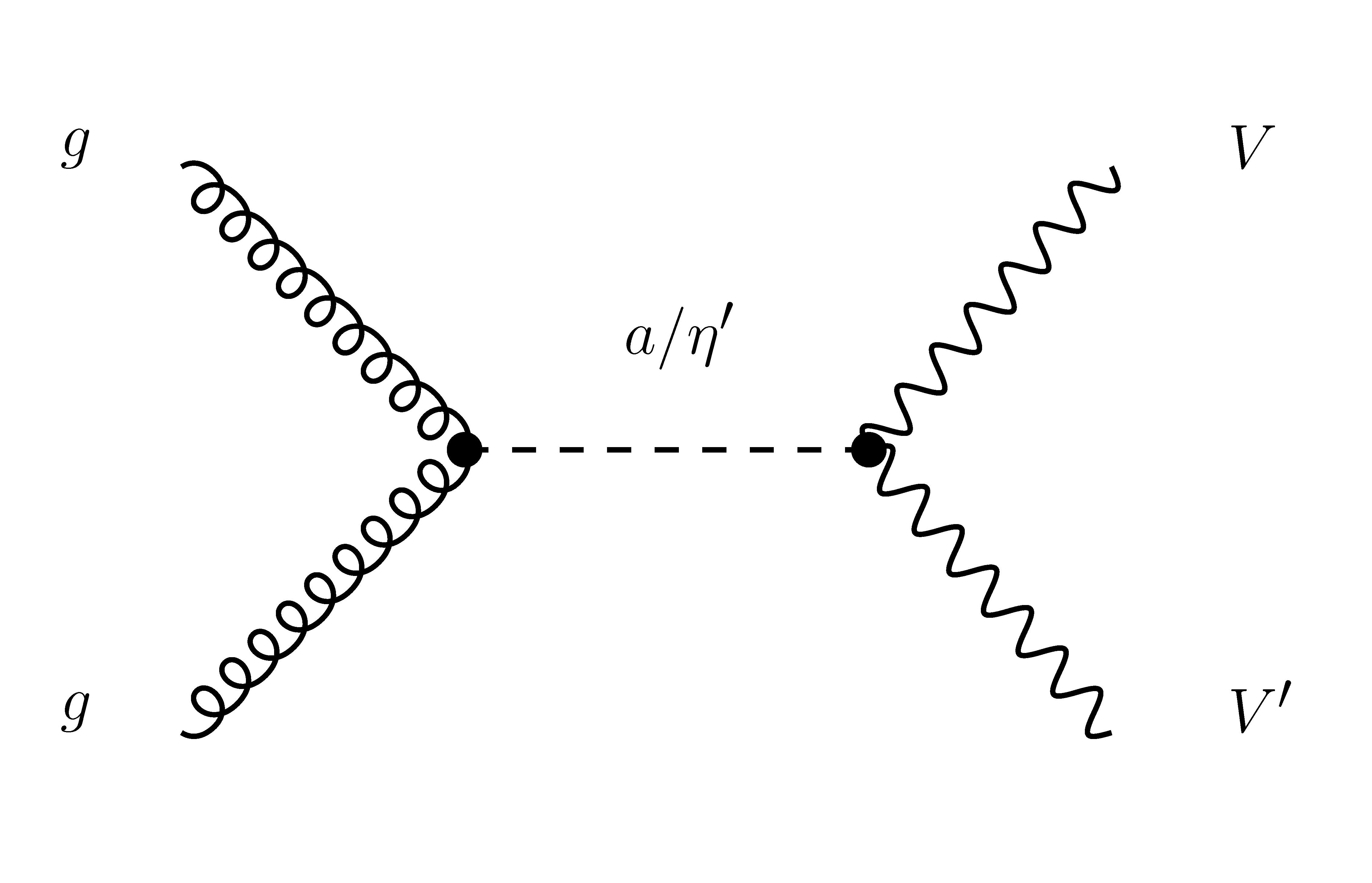}
\caption{The main production mode of the two ALPs is via anomalous gluon fusion. Barring large mixing with other pNGBs, they subsequently decay into a pair of vector bosons with computable Branching Ratios.}
\label{alpfeyn}
\end{center}
\end{figure}

Regardless of their mass, these objects are singly produced mostly by gluons via the anomaly and decay to di-bosons also via the anomaly (Fig.~\ref{alpfeyn}) with calculable branching ratios. This makes them a good window into UV physics since the branching ratios are related to the type of UV d.o.f. of the underlying theory.
It would also be interesting to investigate in detail the mixing of these scalars with the other fields in the EW coset, as done recently in~\cite{Howe:2016mfq} in the context of the model~\cite{Georgi:1985nv}. This could lead to an enhancement in the cross-section for the EW pNGBs.

\subsection*{Acknowledgments}

The author wishes to thank
A.~Belyaev,
G.~Cacciapaglia,
T.~Flacke,
M.~Frigerio,
M.~Golterman,
A.~Hallin,
D.~Karateev,
A.~Padellaro,
C.~Pettersson,
C.~Pica,
F.~Sannino,
Y.~Shamir
and
L.~Vecchi
 for discussion.
 The author would also like to express a special thanks to the Mainz Institute for Theoretical Physics (MITP) for its hospitality and support.

\appendix
\section{All models of partial compositeness satisfying the requirements in the text}

In this appendix we list all models of partial compositeness satisfying the requirements in the text. The main requirements are a simple hypercolor gauge group $\GHC$ and two irreps $\psi$ and $\chi$ giving rise to a custodial EW coset and top partners. In addition, we require the theory to be asymptotically free and of course free of gauge anomalies.

Comparing with~\cite{Ferretti:2013kya} we have removed a few models that do not seem promising. Some are based on spinorial irreps of the orthogonal group for which, as discussed in~\cite{Ferretti:2013kya}, the MAC hypothesis leads to the wrong symmetry breaking pattern. Others are those having baryons of type $\chi\psi\chi$ with $\psi$ in a complex irrep. This leads to top partners in the $({\mathbf{2}}, {\mathbf{1}})$ violating the custodial symmetry~\cite{Agashe:2006at}.

If the di-photon excess~\cite{diphotonATLASCMS} will be confirmed with properties roughly in agreement with the 2015 data, only a fraction of models~\cite{Belyaev:2015hgo} will be able to fit the data. Further restrictions~\cite{Cacciapaglia:2015vrx} could arise from imposing 't~Hooft anomaly matching~\cite{'tHooft:1980xb}.

\begin{table}
\bce
{\footnotesize
  \begin{tabular}{|c|c|c|c|c|}
    \hline
    $\GHC$ & $ \psi$ & $ \chi$ & Restrictions & Top-partners\\
    \hline\hline\noalign{\medskip}
      \multicolumn{1}{c}{} &    \multicolumn{1}{c}{Real} & \multicolumn{1}{c}{Real} & \multicolumn{2}{c}{} \\
      \hline
    $SO(\NHC)$ & $5\times \Sy_2$ & $6\times \Fu$ & $\NHC \geq 55$ & $\chi\psi\chi $ \\
    \hline
    $SO(\NHC)$ & $5\times \Ad$ & $6\times \Fu$ & $\NHC \geq 15$ & $\chi\psi\chi$\\
    \hline
    $SO(\NHC)$ & $5\times \Fu$ & $6\times \Sp$ & $\NHC=7,9$ & $\chi\psi\chi$\\
    \hline
    $SO(\NHC)$ & $5\times \Sp$ & $6\times \Fu$ & $\NHC = 7,9 $ & $\psi\chi\psi $\\
    \hline\hline \noalign{\medskip}
     \multicolumn{1}{c}{} &    \multicolumn{1}{c}{Real} & \multicolumn{1}{c}{Pseudo-Real} & \multicolumn{2}{c}{} \\
     \hline
    $Sp(2 \NHC)$ & $5\times \Ad$ & $6\times \Fu$ & $2 \NHC \geq 12$ &$\chi\psi\chi$\\
    \hline
    $Sp(2 \NHC)$ & $5\times \An_2$ & $6\times \Fu$ & $2 \NHC \geq 4$ &$\chi\psi\chi$\\
    \hline
    $SO(\NHC)$ & $5\times \Fu$ & $6\times \Sp$ & $\NHC=11,13$ &$\chi\psi\chi$ \\
        \hline\hline \noalign{\medskip}
     \multicolumn{1}{c}{} &    \multicolumn{1}{c}{Real} & \multicolumn{1}{c}{Complex} & \multicolumn{2}{c}{} \\
     \hline
    $SU(\NHC)$ & $5\times \An_2$ & $3\times (\Fu, \overline{\Fu})$ & $\NHC = 4$ & $\chi\psi\chi$\\
    \hline
    $SO(\NHC)$ & $5\times \Fu$ & $3\times (\Sp, \overline{\Sp})$ & $\NHC = 10,14$ &$\chi\psi\chi$ \\
    \hline\hline \noalign{\medskip}
     \multicolumn{1}{c}{} &    \multicolumn{1}{c}{Pseudo-Real} & \multicolumn{1}{c}{Real} & \multicolumn{2}{c}{} \\
     \hline
    $Sp(2 \NHC)$ & $4\times \Fu$ & $6\times \An_2$ & $2 \NHC \leq 36$ & $\psi\chi\psi $\\
    \hline
    $SO(\NHC)$ & $4\times \Sp$ & $6\times \Fu$ & $\NHC = 11,13$ & $\psi\chi\psi $\\
    \hline\hline \noalign{\medskip}
     \multicolumn{1}{c}{} &    \multicolumn{1}{c}{Complex} & \multicolumn{1}{c}{Real} & \multicolumn{2}{c}{} \\
     \hline
    $SO(\NHC)$ & $4\times (\Sp,  \overline{\Sp})$ & $6\times \Fu$ & $\NHC = 10$ & $\psi\chi\psi $\\
    \hline
    $SU(\NHC)$ & $4\times(\Fu,  \overline{\Fu})$ & $6\times \An_2$ & $\NHC = 4$ & $\psi\chi\psi $\\
    \hline\hline \noalign{\medskip}
     \multicolumn{1}{c}{} &    \multicolumn{1}{c}{Complex} & \multicolumn{1}{c}{Complex} & \multicolumn{2}{c}{} \\
     \hline
    $SU(\NHC)$ & $4\times (\Fu, \overline{\Fu})$ & $3\times (\An_2, \overline{\An}_2)$ & $\NHC \geq 5$ & $\psi\chi\psi$\\
    \hline
    $SU(\NHC)$ & $4\times (\Fu, \overline{\Fu})$ & $3\times (\Sy_2, \overline{\Sy}_2)$ & $\NHC \geq 5$ & $\psi\chi\psi$\\
    \hline
     $SU(\NHC)$ & $4\times (\An_2, \overline{\An}_2)$ & $3\times (\Fu, \overline{\Fu})$ & $\NHC = 5$ & $\psi\chi\psi$\\
    \hline\hline
  \end{tabular}}
\ece
  \caption{All models obeying the consistency requirements discussed in Appendix~A. This list contains both conformal and confining theories. See text for a discussion of their IR properties.}
  \label{allmodels}
\end{table}

The list of models presented in Table~\ref{allmodels} contains both conformal and confining theories.

It is unfortunately not yet possible to exactly identify the conformal region in non-supersymmetric gauge theories. However, one can use some heuristic arguments to get indications on their behavior and it turns out that most of the models are rather clear-cut cases. Consider for instance the two-loop beta-function
$\beta(\alpha) =   \beta_1 \alpha^2 + \beta_2 \alpha^3$. ($\beta_1 <0$ always.) A formal solution $\alpha^*$ to $\beta(\alpha^*) = 0$ exists for $\beta_2 >  0$ and, if not to large, it can be trusted and the theory can be assumed to be in the conformal regime. If $\beta_2 < 0$ or $\alpha^*$ is out of the perturbative regime, the model is likely to be confining. In between there is a region, difficult to characterize precisely, where the theory is conformal but strongly coupled.

In Table~\ref{noconf} we list the subset of models that are likely to be \emph{outside} of the conformal window. These models also obey 
the heuristic bound  $ 11  C(G) > 4 \left(N_\psi T(\psi) + N_\chi T(\chi)\right) $
proposed in~\cite{CFTbounds} as well as the rigorous bounds from the
$a$-theorem~\cite{atheorem}  $a_{\mathrm{UV}} > a_{\mathrm{IR}}$.
\begin{table}
\bce
{\footnotesize
\def\arraystretch{1.4}%
  \begin{tabular}{|c|c|c|c|c|}
    \hline
    $\GHC$ & $ \psi$ & $ \chi$ & Restrictions & $G/H$ \\
    \hline\hline
    $SO(\NHC)$ & $5\times \Fu$ & $6\times \Sp$ & $\NHC=7,9$ & \multirow{2}{*}{ ${ \frac{SU(5)}{SO(5)}}{ \frac{SU(6)}{SO(6)}} U(1) $}\\
     \cline{1-4}
    $SO(\NHC)$ & $5\times \Sp$ & $6\times \Fu$ & $\NHC = 7,9 $ & \\
    \hline
    $Sp(2 \NHC)$ & $5\times \An_2$ & $6\times \Fu$ & $2 \NHC = 4$ & $ { \frac{SU(5)}{SO(5)}}{ \frac{SU(6)}{Sp(6)}} U(1) $\\
    \hline
    $SU(\NHC)$ & $5\times \An_2$ & $3\times (\Fu, \overline{\Fu})$ & $\NHC = 4$ & \multirow{2}{*}{${ \frac{SU(5)}{SO(5)}}{ \frac{SU(3)\times SU(3)'}{SU(3)_D}} U(1)$ }\\
     \cline{1-4}
    $SO(\NHC)$ & $5\times \Fu$ & $3\times (\Sp, \overline{\Sp})$ & $\NHC = 10$ &\\
     \hline
    $Sp(2 \NHC)$ & $4\times \Fu$ & $6\times \An_2$ & $2 \NHC =4$ & \multirow{2}{*}{${ \frac{SU(4)}{Sp(4)}}{ \frac{SU(6)}{SO(6)}} U(1) $} \\
     \cline{1-4}
    $SO(\NHC)$ & $4\times \Sp$ & $6\times \Fu$ & $\NHC = 11$ &\\
     \hline
     $SO(\NHC)$ & $4\times (\Sp,  \overline{\Sp})$ & $6\times \Fu$ & $\NHC = 10$ & \multirow{2}{*}{${ \frac{SU(4)\times SU(4)'}{SU(4)_D}}{ \frac{SU(6)}{SO(6)}} U(1)$}\\
     \cline{1-4}
    $SU(\NHC)$ & $4\times(\Fu,  \overline{\Fu})$ & $6\times \An_2$ & $\NHC = 4$ &\\
    \hline
    $SU(\NHC)$ & $4\times (\Fu, \overline{\Fu})$ & $3\times (\An_2, \overline{\An}_2)$ & $\NHC = 5, 6$ &
                  ${ \frac{SU(4)\times SU(4)'}{SU(4)_D}}{ \frac{SU(3)\times SU(3)'}{SU(3)_D}} U(1) $ \\
    \hline
  \end{tabular}}
  \ece
  \caption{Subclass of models that is likely to be outside of the conformal window, together with the coset they give rise to after spontaneous symmetry breaking.}
  \label{noconf}
  \end{table}

The use of these models for BSM physics depends on their IR behavior. The simplest application would be to restrict oneself to the models in Table~\ref{noconf}. These models can be easily brought into the conformal window from the strong coupling side by adding additional matter. The most straightforward way of achieving this is to have additional fermions, possibly in the same irreps, with masses at the scale $\Lambda$. In this case one has a concrete way to put the theory in the ``strongest possible'' conformal point where the anomalous dimensions of the top-partners may be large enough. The theory then exits the conformal point at the scale $\Lambda$, where the additional d.o.f. decouple.  However, models outside of this class might still be amenable to other applications and we decided to keep them in the full classification of Table~\ref{allmodels}.

\section{Group theory conventions for the three cosets}

In this appendix we collect the conventions for the explicit constructions of the three EW cosets studied in the text.

\subsection{  \it Notation for the $\Rcoset$ coset}

In this case we realize the Lie algebra of the unbroken group $SO(5)$ as the subset of antisymmetric imaginary generators of $SU(5)$.
This is just a particular choice of basis; a more general way of doing the decomposition is to introduce a symmetric matrix $\delta_0$ and define the broken/unbroken generators as $T\delta_0 \mp \delta_0 T^T = 0 $ respectively. We chose not to do this, and set $\delta_0 = {\mathbf{1}}$ from the onset but comment below on the general form of the pNGB matrix in the general case. The generators of the custodial $SU(2)_L\times SU(2)_R$ are chosen to be

{\footnotesize
\beqs
      &&T_L^1 = \left(
\begin{array}{ccccc}
 0 & 0 & 0 & -\frac{i}{2} & 0 \\
 0 & 0 & -\frac{i}{2} & 0 & 0 \\
 0 & \frac{i}{2} & 0 & 0 & 0 \\
 \frac{i}{2} & 0 & 0 & 0 & 0 \\
 0 & 0 & 0 & 0 & 0
\end{array}\right),\;
   T_L^2 = \left(
\begin{array}{ccccc}
 0 & 0 & \frac{i}{2} & 0 & 0 \\
 0 & 0 & 0 & -\frac{i}{2} & 0 \\
 -\frac{i}{2} & 0 & 0 & 0 & 0 \\
 0 & \frac{i}{2} & 0 & 0 & 0 \\
 0 & 0 & 0 & 0 & 0
\end{array}
\right),\;
T_L^3 = \left(
      \begin{array}{ccccc}
 0 & -\frac{i}{2} & 0 & 0 & 0 \\
 \frac{i}{2} & 0 & 0 & 0 & 0 \\
 0 & 0 & 0 & -\frac{i}{2} & 0 \\
 0 & 0 & \frac{i}{2} & 0 & 0 \\
 0 & 0 & 0 & 0 & 0
\end{array}
\right),\;
\nn\\
     && T_R^1 = \left(
\begin{array}{ccccc}
 0 & 0 & 0 & \frac{i}{2} & 0 \\
 0 & 0 & -\frac{i}{2} & 0 & 0 \\
 0 & \frac{i}{2} & 0 & 0 & 0 \\
 -\frac{i}{2} & 0 & 0 & 0 & 0 \\
 0 & 0 & 0 & 0 & 0
\end{array}
\right), \;
     T_R^2 = \left(
\begin{array}{ccccc}
 0 & 0 & \frac{i}{2} & 0 & 0 \\
 0 & 0 & 0 & \frac{i}{2} & 0 \\
 -\frac{i}{2} & 0 & 0 & 0 & 0 \\
 0 & -\frac{i}{2} & 0 & 0 & 0 \\
 0 & 0 & 0 & 0 & 0
\end{array}
\right), \;
      T_R^3 = \left(
\begin{array}{ccccc}
 0 & -\frac{i}{2} & 0 & 0 & 0 \\
 \frac{i}{2} & 0 & 0 & 0 & 0 \\
 0 & 0 & 0 & \frac{i}{2} & 0 \\
 0 & 0 & -\frac{i}{2} & 0 & 0 \\
 0 & 0 & 0 & 0 & 0
\end{array}
\right).
\eeqs}

The broken generators are the real symmetric traceless generators of $SU(5)$. We write the pNGBs as
{\footnotesize \beq
    H =
        \begin{pmatrix}
          0 & 0 & 0 & 0 & -i H_+/\sqrt{2} \\
          0 & 0 & 0 & 0 & H_+/\sqrt{2} \\
          0 & 0 & 0 & 0 & i H_0/\sqrt{2} \\
          0 & 0 & 0 & 0 & H_0/\sqrt{2} \\
          -i H_+/\sqrt{2} & H_+/\sqrt{2} & i H_0/\sqrt{2} & H_0/\sqrt{2} & 0\\
        \end{pmatrix}
        \eeq
        \beq
   \Phi_0 = \begin{pmatrix}
                \phi_0^0/\sqrt{2} & 0 & i(\phi_0^- - \phi_0^+)/2 & (\phi_0^- + \phi_0^+)/2 & 0 \\
                0 & \phi_0^0/\sqrt{2} & (\phi_0^- + \phi_0^+)/2  & -i(\phi_0^- - \phi_0^+)/2  & 0 \\
                i(\phi_0^- - \phi_0^+)/2 & (\phi_0^- + \phi_0^+)/2 & -\phi_0^0/\sqrt{2} & 0 & 0 \\
                (\phi_0^- + \phi_0^+)/2 & -i(\phi_0^- - \phi_0^+)/2 & 0 & -\phi_0^0/\sqrt{2} & 0 \\
                0 & 0 & 0 & 0 & 0
              \end{pmatrix}
         \eeq
        \beq
    \Phi_+ = \begin{pmatrix}
                   \phi_+^+/\sqrt{2} & i\phi_+^+/\sqrt{2} & i\phi_+^0/2 & \phi_+^0/2 & 0 \\
                   i\phi_+^+/\sqrt{2} & -\phi_+^+/\sqrt{2} & -\phi_+^0/2 & i\phi_+^0/2 & 0 \\
                   i\phi_+^0/2 & -\phi_+^0/2 & \phi_+^-/\sqrt{2} & -i\phi_+^-/\sqrt{2}  & 0 \\
                   \phi_+^0/2 &  i\phi_+^0/2 & -i\phi_+^-/\sqrt{2}  & -\phi_+^-/\sqrt{2}  & 0 \\
                   0 & 0 & 0 & 0 & 0 \\
                 \end{pmatrix}
         \eeq
        \beq
    E  = \begin{pmatrix}
                   \frac{\eta }{\sqrt{10}} & 0 & 0 & 0 & 0 \\
                    0 & \frac{\eta }{\sqrt{10}} & 0 & 0 & 0 \\
                    0 & 0 & \frac{\eta }{\sqrt{10}} & 0 & 0 \\
                    0 & 0 & 0 & \frac{\eta}{\sqrt{10}} & 0 \\
                    0 & 0 & 0 & 0 & -2 \sqrt{\frac{2}{5}} \eta
           \end{pmatrix}
\eeq}
In this way with our conventions $\phi_m^{n*} = \phi_{-m}^{-n}$ the full matrix of pNGBs is real symmetric::
\beq
    \Pi = H + H^\dagger + \Phi_0 + \Phi_+ + \Phi_+^\dagger + E.
\eeq

The vacuum misalignment is described by the following unitary matrix obtained by exponentiating (half of) the Higgs v.e.v.
\beq
      \Omega = \left(
        \begin{array}{ccccc}
        1 & 0 & 0 & 0 & 0 \\
        0 & 1 & 0 & 0 & 0 \\
        0 & 0 & 1 & 0 & 0 \\
        0 & 0 & 0 & \cos\zeta & i \sin\zeta \\
        0 & 0 & 0 & i \sin\zeta & \cos\zeta
        \end{array} \right).
\eeq
$\Omega$ preserves the custodial symmetry $SU(2)_D$ generated by $T_L^i + T_R^i$ and we write the non-linear realization of the pNGBs as a symmetric and unitary matrix
\beq
    U = \Omega \exp(2 i \Pi/f) \Omega^T.
\eeq
All the fields in $\Pi$ have zero v.e.v. and in the unitary gauge $H_+ = 0$ and $H_0 = h/\sqrt{2}$.

Notice that with our choice of basis, $\Omega = \Omega^T$. Had we chosen a more general $\delta_0$, we would have obtained
\beq
    U = \Omega \exp(2 i \Pi/f)\delta_0 \Omega^T = \Omega \exp(2 i \Pi/f) \Omega \delta_0 \equiv \Sigma \delta_0
\eeq
where the last identity defines $\Sigma$. The matrix $\Sigma$ has the advantage of making some formulas look more uniform in all three cases but the disadvantage of not transforming uniformly under $SU(5)$ and we chose not to use it.
The covariant derivative is
\beq
     D_\mu U = \partial_\mu U - i g W_\mu^i(T_L^i U + U T_L^{iT}) - i g' B_\mu (T_R^3 U + U T_R^{3T}) \label{covderSO}
\eeq
and in our convention can be written in terms of commutators. Finally, the kinetic term is
\beq
      L_{\mathrm{kin}} = \frac{f^2}{16}\int \tr(D_\mu U^\dagger D^\mu U).
\eeq
\subsection{  \it Notation for the $\PRcoset$ coset}

We pick the symplectic matrix
\beq
          \epsilon_0 = \left(
        \begin{array}{cccc}
        0 & 1 & 0 & 0 \\
       -1 & 0 & 0 & 0 \\
        0 & 0 & 0 & -1 \\
        0 & 0 & 1 & 0
        \end{array} \right).
\eeq
The unbroken generators satisfy $ T^i \epsilon_0 + \epsilon_0 T^{iT} = 0$. In particular, the generators of $SU(2)_L\times SU(2)_R$ are chosen to be

{\footnotesize
\beqs
      &&T_L^1 =\left(
\begin{array}{cccc}
 0 & \frac{1}{2} & 0 & 0 \\
 \frac{1}{2} & 0 & 0 & 0 \\
 0 & 0 & 0 & 0 \\
 0 & 0 & 0 & 0
\end{array}
\right),\;
      T_L^2 = \left(
\begin{array}{cccc}
 0 & -\frac{i}{2} & 0 & 0 \\
 \frac{i}{2} & 0 & 0 & 0 \\
 0 & 0 & 0 & 0 \\
 0 & 0 & 0 & 0
\end{array}
\right),\;
     T_L^3 =\left(
\begin{array}{cccc}
 \frac{1}{2} & 0 & 0 & 0 \\
 0 & -\frac{1}{2} & 0 & 0 \\
 0 & 0 & 0 & 0 \\
 0 & 0 & 0 & 0
\end{array}
\right),\;
\nn\\
     && T_R^1 = \left(
\begin{array}{cccc}
 0 & 0 & 0 & 0 \\
 0 & 0 & 0 & 0 \\
 0 & 0 & 0 & \frac{1}{2} \\
 0 & 0 & \frac{1}{2} & 0
\end{array}
\right), \;
     T_R^2 = \left(
\begin{array}{cccc}
 0 & 0 & 0 & 0 \\
 0 & 0 & 0 & 0 \\
 0 & 0 & 0 & -\frac{i}{2} \\
 0 & 0 & \frac{i}{2} & 0
\end{array}
\right),\;
      T_R^3 = \left(
\begin{array}{cccc}
 0 & 0 & 0 & 0 \\
 0 & 0 & 0 & 0 \\
 0 & 0 & \frac{1}{2} & 0 \\
 0 & 0 & 0 & -\frac{1}{2}
\end{array}
\right)
\eeqs}
and the pNGBs can be represented as
{\footnotesize \beqs
H = \left(
\begin{array}{cccc}
 0 & 0 &
   \frac{i
   {H_0}^*}{2} &
   \frac{i H_+}{2} \\
 0 & 0 &
   -\frac{i {H_+}^*}{2} &
   \frac{i H_0}{2} \\
   0&0&0&0\\
   0&0&0&0
\end{array}
\right),\; E =  \left(
\begin{array}{cccc}
 \frac{\eta }{2 \sqrt{2}} & 0 & 0 & 0 \\
   0 & \frac{\eta }{2 \sqrt{2}} &0 & 0 \\
   0 & 0 & -\frac{\eta }{2 \sqrt{2}} & 0\\
   0 & 0 & 0 & -\frac{\eta }{2 \sqrt{2}}
\end{array}
\right)
\eeqs}
\beq
      \Pi = H + H^\dagger + E.
\eeq
Notice that $\Pi\epsilon_0 - \epsilon_0 \Pi^T = 0$.

The matrix describing vacuum misalignment and preserving the custodial symmetry is
\beq
      \Omega(\theta) = \left(
\begin{array}{cccc}
 \cos\frac{\zeta
   }{2} & 0 & -\sin
   \frac{\zeta}{2} & 0 \\
 0 & \cos \frac{\zeta }{2}& 0 & -\sin
   \frac{\zeta}{2} \\
 \sin\frac{\zeta}{2} & 0 & \cos
   \frac{\zeta }{2} & 0 \\
 0 & \sin\frac{\zeta}{2} & 0 & \cos
   \frac{\zeta}{2}
\end{array}
\right),
\eeq
in terms of which the non-linear realization can be expressed as an anti-symmetric and unitary matrix
\beq
     U = \Omega \exp(2 \sqrt{2} i \Pi/f)\epsilon_0 \Omega^T.
\eeq
Also in this case, the fields in $\Pi$ have zero v.e.v. and in the unitary gauge $H_+ = 0$ and $H_0 = h/\sqrt{2}$.
The covariant derivative reads as in the previous case (\ref{covderSO}) but the kinetic term is normalized differently
\beq
      L_{\mathrm{kin}} = \frac{f^2}{8}\int \tr(D_\mu U^\dagger D^\mu U).
\eeq

Even in this case one has the option of using $\epsilon_0 \Omega^T = \Omega \epsilon_0$ and of introducing a matrix $\Sigma$ through the identity $U = \Sigma \epsilon_0$ in an analogous way as for the previous coset, but we do not use it for the same reasons as above.

\subsection{  \it Notation for the $\Ccoset$ coset}
In this case, the $SU(2)_L\times SU(2)_R$ subgroup is embedded in the unbroken $SU(4)_D$ by choosing
{\footnotesize
\beqs
      &&T_L^1 =\left(
\begin{array}{cccc}
 0 & \frac{1}{2} & 0 & 0 \\
 \frac{1}{2} & 0 & 0 & 0 \\
 0 & 0 & 0 & 0 \\
 0 & 0 & 0 & 0
\end{array}
\right) ,\;
      T_L^2 = \left(
\begin{array}{cccc}
 0 & -\frac{i}{2} & 0 & 0 \\
 \frac{i}{2} & 0 & 0 & 0 \\
 0 & 0 & 0 & 0 \\
 0 & 0 & 0 & 0
\end{array}
\right),\;
     T_L^3 =\left(
\begin{array}{cccc}
 \frac{1}{2} & 0 & 0 & 0 \\
 0 & -\frac{1}{2} & 0 & 0 \\
 0 & 0 & 0 & 0 \\
 0 & 0 & 0 & 0
\end{array}
\right),\;
\nn\\
     && T_R^1 = \left(
\begin{array}{cccc}
 0 & 0 & 0 & 0 \\
 0 & 0 & 0 & 0 \\
 0 & 0 & 0 & \frac{1}{2} \\
 0 & 0 & \frac{1}{2} & 0
\end{array}
\right),\;
     T_R^2 =\left(
\begin{array}{cccc}
 0 & 0 & 0 & 0 \\
 0 & 0 & 0 & 0 \\
 0 & 0 & 0 & -\frac{i}{2} \\
 0 & 0 & \frac{i}{2} & 0
\end{array}
\right),\;
      T_R^3 = \left(
\begin{array}{cccc}
 0 & 0 & 0 & 0 \\
 0 & 0 & 0 & 0 \\
 0 & 0 & \frac{1}{2} & 0 \\
 0 & 0 & 0 & -\frac{1}{2}
\end{array}
\right).
\eeqs}

The pNGBs are parameterized as follows
{\footnotesize \beqs
&&H = \left(
\begin{array}{cccc}
 0 & 0 & -\frac{1}{2} i \left( {H}^*_0+i {H'}^*_0\right) & -\frac{1}{2} i \left(H_++i H'_+\right) \\
 0 & 0 & \frac{1}{2} i \left({H}^*_+ +i {H'}^*_+\right) & -\frac{1}{2} i \left(H_0+i H'_0\right) \\
 0 & 0 & 0 & 0 \\
 0 & 0 & 0 & 0
\end{array}
\right),\;
E =
\left(
\begin{array}{cccc}
 \frac{1}{2}\frac{\eta}{\sqrt{2}} & 0 & 0 & 0 \\
 0 & \frac{1}{2} \frac{\eta}{\sqrt{2}} & 0 & 0 \\
 0 & 0 & -\frac{1}{2}\frac{\eta}{\sqrt{2}} &  0 \\
 0 & 0 & 0 & -\frac{1}{2}\frac{\eta}{\sqrt{2}}
\end{array}
\right)
\nn \\
&&\Phi = \left(
\begin{array}{cccc}
 \frac{1}{2}\phi_0 &
   \frac{1}{\sqrt 2} \phi_+ & 0 & 0 \\
   \frac{1}{\sqrt 2}\phi_-& -\frac{1}{2} \phi_0 & 0 & 0 \\
   0 & 0 & 0 & \\
   0 & 0 & 0 &
\end{array}
\right), \;
N = \left(
\begin{array}{cccc}
  0 & 0 & 0 & 0 \\
  0 & 0 & 0 & 0 \\
  0 & 0 & \frac{1}{2} N_0 & \frac{1}{\sqrt 2} N_+\\
  0 & 0 & \frac{1}{\sqrt 2} N_- & -\frac{1}{2} N_0
\end{array}
\right),
\eeqs}
where $\phi_+^* = \phi_-$ and $N_+^* = N_-$. In the unitary gauge we have as usual $H_+=0$, $H_0=h/\sqrt{2}$, having chosen to rotate the v.e.v. into the first of the two doublets. The neutral component of the second doublet is thus physical and can be written as a CP even plus CP odd part:
$H'_0 = (h' + i A')/\sqrt{2}$. Finally we set
\beq
     \Pi = H + H^\dagger + N + \Phi +E
\eeq
and, for the vacuum misalignment matrix we obtain
\beq
      \Omega(\theta) = \left(
\begin{array}{cccc}
 \cos \frac{\zeta
   }{2} & 0 & \sin
   \frac{\zeta
   }{2} & 0 \\
 0 & \cos  \frac{\zeta
   }{2}  & 0 & \sin
   \frac{\zeta
   }{2} \\
 -\sin \frac{\zeta
   }{2}& 0 & \cos
   \frac{\zeta
   }{2} & 0 \\
 0 & -\sin\frac{\zeta
   }{2}& 0 & \cos
   \frac{\zeta   }{2}
\end{array}
\right).
\eeq
In this case, the non linear realization of the pNGBs is given by the unitary matrix
\beq
       U = \Omega \exp(2 \sqrt{2} i \Pi/f) \Omega.
\eeq
The covariant derivative is obtained by the usual commutator
\beq
     D_\mu U = \partial_\mu U - i g W_\mu^i [T_L^i, U] - i g' B_\mu [T_R^3, U]
\eeq
and the kinetic term is normalized as
\beq
      L_{\mathrm{kin}} = \frac{f^2}{8}\int \tr(D_\mu U^\dagger D^\mu U).
\eeq

\section{Additional three and four bosons couplings for the models in the text.}

Additional (i.e. other than those involving $h$) quartic couplings for $\Ccoset$ are shown in~(\ref{SUquartic}) (in agreement with \cite{Ma:2015gra}).

\begin{align}
    {\mathcal{L}} \supset
    & e^2 A_\mu A^\mu\Big(N_- N_+ + \phi_- \phi_+ + H'_- H'_+\Big)\nn\\
    & +\frac{e^2}{\sw \cw}Z_\mu A^\mu\Big((\cww-\cz)N_- N_+ + (\cww+\cz)\phi_-\phi_+ + \cww H'_- H'_+\Big)\nn\\
    & +\frac{e^2}{16 \sw^2 \cw^2}Z_\mu Z^\mu\Big( 2 \cz^2 h' h' + 2(\cwwww +3\cz^2 - 4\cww\cz) N_- N_+ + 2(\cwwww + 3\cz^2 + 4\cww\cz) \phi_- \phi_+
    \nn\\ & ~~~~~~~~~~- 2 \sz^2 N_+ \phi_- - 2 \sz^2 \phi_+ N_-
    - \sz^2 N_0 N_0 + 2\sz^2 N_0 \phi_0  - \sz^2 \phi_0 \phi_0 + 2\czz A' A' - 2 \sz^2 \eta \eta + 4 \cww^2 H'_+ H'_-\Big)\nn\\
     &+\frac{e^2}{8 \sw^2}W^-_\mu W^{+\mu}\Big( 2 \cz^2 h' h' - 4(1-\cz)\cz N_- N_+ + 4(1+\cz)\cz \phi_- \phi_+ +(1-\cz)(1-3\cz)N_0 N_0
     \nn\\ &~~~~~~~~~~- 2 \sz^2 N_0\phi_0 +(1+\cz)(1+3\cz)\phi_0\phi_0 + 2 A' A' - 2 \sz^2 \eta\eta + 4\cz^2 H'_+ H'_-\Big)\nn\\
     &+\frac{e^2}{2 \sw}A_\mu W^{-\mu}\Big(-(1-\cz)N_0 N_+ - (1+\cz)\phi_0 \phi_+ - i A' H'_+ + \cz h' H'_+\Big)\nn\\
     & -\frac{e^2}{4 \sw^2 \cw}Z_\mu W^{-\mu}\Big((1-\cz)(\cww-\cz)N_0 N_+ -  \sz^2 \phi_0 N_+ - \sz^2 \phi_+ N_0
     \nn \\ & ~~~~~~~~~~+(1+\cz)(\cww+\cz)\phi_+ \phi_0 +  i (\cww -\czz)A' H'_+ + 2\cz \sw^2 h'H'_+\Big)\nn\\
     & -\frac{e^2}{8 \sw^2 }W^-_\mu W^{-\mu}\Big((1-\cz)^2 N_+ N_+ -2 \sz^2 N_+ \phi_+ + (1+\cz)^2 \phi_+ \phi_+ - 2 \sz^2 H_+ H_+\Big)\nn\\
     &+ \mbox{hermitian conjugates of the terms involving $A W^-$, $Z W^-$, $W^- W^-$.}
     \label{SUquartic}
\end{align}

The anomalous cubic couplings for $\Rcoset$ are shown in Table~\ref{SU5WZW}.  Each coupling should be multiplied by $ e^2\dim(\psi)/(48\pi^2 f)$.

Finally, we present in eq.~(\ref{SOquartic}) the additional (i.e. other than those involving $h$) quartic couplings for $\Rcoset$.

Of course, the generation of masses by the potential introduces a mixing between these gauge eigenstates. This depends on the specific nature of the mass matrix and in many cases it could be handled by the mass insertion approximation. Throughout the paper we work with gauge eigenstates. Also note that one could use the  Clebsch - Gordan coefficients to express the gauge eigenstates as eigenstates of the diagonal custodial symmetry group $SU(2)_D \subset SU(2)_L\times SU(2)_R$ as done in \cite{Dugan:1984hq}.
An even deeper difference with the model in~\cite{Dugan:1984hq} is that they used an additional $U(1)$ gauge field to induce vacuum-misalignment instead of top coupling.
\begin{table}[h]
\def\arraystretch{2.}%
\begin{tabular}{|c|c||c|c||c|c|}
  \hline
  Fields & Couplings & Fields & Couplings & Fields & Couplings\\
  \hline
  $\eta F_{\mu\nu}\tilde F^{\mu\nu}$ & $3 \sqrt{\frac{2}{5}}$  &   $\phi_0^0 F_{\mu\nu}\tilde Z^{\mu\nu}$ & $6 \sqrt{2} \cww/\sww$  &
  $\phi_+^0 F_{\mu\nu}\tilde W^{-\mu\nu}$ & $\frac{3 i (\cz-1) }{\sqrt{2}\sw}$  \\
  \hline
  $\eta F_{\mu\nu}\tilde Z^{\mu\nu}$ & $6 \sqrt{\frac{2}{5}}  \cww/\sww$ & $\phi_0^0 Z_{\mu\nu}\tilde Z^{\mu\nu}$ & $\frac{
   (-5 \czz+6 \cwwww-1)}{2 \sqrt{2}\sww^2}$ & $\phi_+^0 Z_{\mu\nu}\tilde W^{-\mu\nu}$ & $-\frac{i(1-\cz)
  (2 \cz +3 \cww-1)}{2\sqrt{2}\sw^2 \cw}$  \\
  \hline
  $\eta Z_{\mu\nu}\tilde Z^{\mu\nu}$ & $\frac{3  (3 \cz^2+\cwwww)}{4 \sqrt{10}\cw^2 \sw^2}$ &
    $\phi_0^0 W^+_{\mu\nu}\tilde W^{-\mu\nu}$ & $\frac{\sz^2 }{2 \sqrt{2}\sw^2}$ &  $\phi_0^+ F_{\mu\nu}\tilde W^{-\mu\nu}$ &
    $\frac{3 i (\cz+1) }{\sqrt{2}\sw}$
  \\
  \hline
  $\eta W^+_{\mu\nu}\tilde W^{-\mu\nu}$ & $\frac{3 (3 \czz+5) }{4 \sqrt{10}\sw^2}$ &
    $\phi_+^- Z_{\mu\nu}\tilde Z^{\mu\nu}$ & $\frac{\sz^2}{\sqrt{2}\sww^2}$ &
    $\phi_0^+ Z_{\mu\nu}\tilde W^{-\mu\nu}$ & $-\frac{i (1+\cz)
   (2 \cz -3 \cww+1)}{2 \sqrt{2}\sw^2 \cw}$  \\
  \hline
  $\phi_0^0 F_{\mu\nu}\tilde F^{\mu\nu}$ & $3 \sqrt{2}$  &   $\phi_+^- W^+_{\mu\nu}\tilde W^{-\mu\nu}$ & $\frac{3 \sz^2}
  {2 \sqrt{2}\sw^2}$  &   $\phi_+^+ W^-_{\mu\nu}\tilde W^{-\mu\nu}$ & $-\frac{\sz^2}{\sqrt{2}\sw^2}$ \\
  \hline
\end{tabular}
\caption{Anomalous couplings for $\Rcoset$, to be multiplied by $ e^2\dim(\psi)/(48\pi^2 f)$.}
\label{SU5WZW}
\end{table}

\begin{align}
    {\mathcal{L}} \supset& e^2 A_\mu A^\mu \Big(\phi_-^0\phi_+^0 + \phi_0^-\phi_0^+ + 4 \phi_-^-\phi_+^+ \Big) \nn\\
    &+ \frac{e^2}{\sw \cw}Z_\mu A^\mu \Big((\cww - \cz)\phi_-^0\phi_+^0 +(\cww + \cz)\phi_0^-\phi_0^+  + 4 \cww \phi_-^-\phi_+^+ \Big)\nn\\
    &+ \frac{e^2}{16 \sw^2 \cw^2}Z_\mu Z^\mu \Big(-5 \sz^2 \eta \eta+ 2\sqrt{5}\sz^2 \eta \phi_0^0 - 2\sqrt{5}\sz^2 \eta \phi_-^+
     - 2\sqrt{5}\sz^2 \eta \phi_+^-  + 2\sz^2 \phi_0^0\phi_-^+  + 2\sz^2 \phi_0^0\phi_+^- - \sz^2 \phi_0^0 \phi_0^0
    \nn\\ &~~~~~~~~~~+ (6 + 10\czz) \phi_-^+\phi_+^- +
    (2 \cwwww - 8 \cz \cww + 6 \cz^2)\phi^0_+ \phi^0_- + 2 \sz^2 \phi^0_+ \phi^-_0
    \nn\\ &~~~~~~~~~~+ (2 \cwwww + 8 \cz \cww + 6 \cz^2)\phi^+_0 \phi^-_0  + 2 \sz^2 \phi^0_- \phi^+_0 + 16 \cww^2 \phi_+^+\phi_-^-\Big)\nn\\
    &+\frac{e^2}{16 \sw^2}W^-_\mu W^{+\mu}\Big (-10 \sz^2 \eta\eta - 4\sqrt{5} \sz^2 \eta\phi_0^0 + (11+5\czz)\phi_0^0\phi_0^0 +
     (12 + 4\czz)\phi_-^+\phi_+^- -8 \sz^2 \phi_0^0 \phi_-^+
    \nn\\ &~~~~~~~~~~-8 \sz^2 \phi_0^0 \phi_+^- - 8\sz^2 \phi^+_0 \phi^0_- +
      (16 +8\cz + 8 \czz)\phi^0_+\phi^0_- - 8\sz^2 \phi^0_+ \phi^-_0 + (16 - 8\cz + 8 \czz)\phi^-_0\phi^+_0 + 16 \cz^2 \phi_-^- \phi_+^+ \Big)\nn\\
    &-\frac{i e^2}{2 \sw}A_\mu W^{-\mu}\Big((1+\cz)\phi^+_- \phi^0_+ -
    (1-\cz)\phi^0_0 \phi^0_+ -(1-\cz)\phi^-_+ \phi^+_0 +(1+\cz)\phi^0_0 \phi^+_0
     \nn\\&~~~~~~~~~~- 3(1 + \cz)\phi_+^+\phi^0_- +3 (1 -\cz) \phi_+^+\phi^-_0\Big)\nn\\
    & -\frac{i e^2}{4 \sw^2 \cw}Z_\mu W^{-\mu}(\sqrt{5}\sz^2 \eta\phi^0_+ + (1+\cz)(1+\cww-4 \cz)\phi^+_- \phi^0_+ + (1-\cz)(\cz - \cww)\phi^0_0 \phi^0_+
   \nn\\ &~~~~~~~~~~ - \sqrt{5}\sz^2 \eta\phi^+_0 + (1+\cz)(\cww + \cz)\phi^+_0\phi^0_0 - (1-\cz)(1+\cww+4\cz)\phi^+_0\phi^-_+\nn\\
   &~~~~~~~~~~ - (1+\cz)(1+3\cww - 2\cz) \phi_+^+ \phi^0_- + (1-\cz)(1+3\cww + 2\cz) \phi_+^+ \phi_0^- \Big)\nn\\
    & + \frac{e^2}{8 \sw^2 }W^-_\mu W^{-\mu}\Big ( (1-\cz)^2\phi^0_+ \phi^0_+ + (1+\cz)^2\phi^+_0 \phi^+_0 - 6\sz^2 \phi^+_0 \phi^0_+
    -2\sqrt{5}\sz^2 \eta\phi_+^+  \nn\\ &~~~~~~~~~~ + 2(1+\cz)^2\phi^+_-\phi_+^+ + 2(1-\cz)^2\phi^-_+\phi_+^+ -6 \sz^2 \phi^0_0\phi_+^+ \Big)\nn\\
     &+ \mbox{hermitian conjugates of the terms involving $A W^-$, $Z W^-$, $W^- W^-$.}
    \label{SOquartic}
\end{align}



\begin{thebibliography}{99}

\bibitem{BEH}
  F.~Englert and R.~Brout,
  Phys.\ Rev.\ Lett.\  {\bf 13} (1964) 321.
%
$\bullet$
    P.~W.~Higgs,
  Phys.\ Rev.\ Lett.\  {\bf 13} (1964) 508.
%
$\bullet$
  G.~S.~Guralnik, C.~R.~Hagen and T.~W.~B.~Kibble,
  Phys.\ Rev.\ Lett.\  {\bf 13} (1964) 585.

\bibitem{Weinberg:1967tq}
  S.~Weinberg,
  Phys.\ Rev.\ Lett.\  {\bf 19} (1967) 1264.

\bibitem{Kaplan:1983fs}
  D.~B.~Kaplan and H.~Georgi,
  Phys.\ Lett.\ B {\bf 136} (1984) 183.

\bibitem{Kaplan:1991dc}
  D.~B.~Kaplan,
  Nucl.\ Phys.\ B {\bf 365} (1991) 259.

\bibitem{CCWZ}
  S.~R.~Coleman, J.~Wess and B.~Zumino,
  Phys.\ Rev.\  {\bf 177} (1969) 2239.
%
$\bullet$
  C.~G.~Callan, Jr., S.~R.~Coleman, J.~Wess and B.~Zumino,
  Phys.\ Rev.\  {\bf 177} (1969) 2247.

\bibitem{reviews}
  R.~Contino,
  arXiv:1005.4269 [hep-ph].
%
$\bullet$
   G.~Panico and A.~Wulzer,
  Lect.\ Notes Phys.\  {\bf 913} (2016) pp.1
  [arXiv:1506.01961 [hep-ph]].

\bibitem{Barnard:2013zea}
  J.~Barnard, T.~Gherghetta and T.~S.~Ray,
  JHEP {\bf 1402} (2014) 002
  [arXiv:1311.6562 [hep-ph]].

\bibitem{Ferretti:2014qta}
  G.~Ferretti,
  JHEP {\bf 1406} (2014) 142
  [arXiv:1404.7137 [hep-ph]].

\bibitem{Vecchi:2015fma}
  L.~Vecchi,
  arXiv:1506.00623 [hep-ph].

\bibitem{Ferretti:2013kya}
  G.~Ferretti and D.~Karateev,
  JHEP {\bf 1403} (2014) 077
  [arXiv:1312.5330 [hep-ph] (V2)].

\bibitem{SUSYcomp}
  F.~Caracciolo, A.~Parolini and M.~Serone,
  JHEP {\bf 1302} (2013) 066
  [arXiv:1211.7290 [hep-ph]].
%
$\bullet$~D.~Marzocca, A.~Parolini and M.~Serone,
  JHEP {\bf 1403} (2014) 099
  [arXiv:1312.5664 [hep-ph]].

\bibitem{otherave}
  R.~Nevzorov and A.~W.~Thomas,
  Phys.\ Rev.\ D {\bf 92} (2015) 075007
  [arXiv:1507.02101 [hep-ph]].
%
  $\bullet$~N.~Bizot and M.~Frigerio,
  JHEP {\bf 1601} (2016) 036
  [arXiv:1508.01645 [hep-ph]].
%
   $\bullet$~V.~Sanz and J.~Setford,
  JHEP {\bf 1512} (2015) 154
  [arXiv:1508.06133 [hep-ph]].

\bibitem{AdditionalPheno}
  A.~De Simone, O.~Matsedonskyi, R.~Rattazzi and A.~Wulzer,
  JHEP {\bf 1304} (2013) 004
  [arXiv:1211.5663 [hep-ph]].
%
$\bullet$
  A.~Thamm, R.~Torre and A.~Wulzer,
  JHEP {\bf 1507} (2015) 100
  [arXiv:1502.01701 [hep-ph]].
%
$\bullet$
  J.~Barnard and M.~White,
  JHEP {\bf 1510} (2015) 072
  [arXiv:1507.02332 [hep-ph]].
%
$\bullet$
  D.~Croon, B.~M.~Dillon, S.~J.~Huber and V.~Sanz,
  arXiv:1510.08482 [hep-ph].
%
$\bullet$
  M.~J.~Schlaffer,
  DESY-THESIS-2015-036.
%
$\bullet$
  C.~Englert, R.~Rosenfeld, M.~Spannowsky and A.~Tonero,
  arXiv:1603.05304 [hep-ph].

\bibitem{Luty:2004ye}
  M.~A.~Luty and T.~Okui,
  JHEP {\bf 0609} (2006) 070
  [hep-ph/0409274].

\bibitem{Strassler:2003ht}
  M.~J.~Strassler,
  hep-th/0309122.

\bibitem{confo}
  R.~Rattazzi, V.~S.~Rychkov, E.~Tonni and A.~Vichi,
  JHEP {\bf 0812} (2008) 031
  [arXiv:0807.0004 [hep-th]].
%
  V.~S.~Rychkov and A.~Vichi,
  Phys.\ Rev.\ D {\bf 80} (2009) 045006
  [arXiv:0905.2211 [hep-th]].

\bibitem{VafaWitten}
  C.~Vafa and E.~Witten,
  Nucl.\ Phys.\ B {\bf 234} (1984) 173.

\bibitem{Matsedonskyi:2014iha}
  O.~Matsedonskyi,
  JHEP {\bf 1502} (2015) 154
  [arXiv:1411.4638 [hep-ph]].

\bibitem{Cacciapaglia:2015dsa}
  G.~Cacciapaglia, H.~Cai, T.~Flacke, S.~J.~Lee, A.~Parolini and H.~Serodio,
  JHEP {\bf 1506} (2015) 085
  [arXiv:1501.03818 [hep-ph]].

\bibitem{Panico:2016ull}
  G.~Panico and A.~Pomarol,
  arXiv:1603.06609 [hep-ph].

\bibitem{Ma:2015gra}
  T.~Ma and G.~Cacciapaglia,
  JHEP {\bf 1603} (2016) 211
  [arXiv:1508.07014 [hep-ph]].

\bibitem{diphotonATLASCMS}
  The ATLAS collaboration,
  ATLAS-CONF-2015-081.
%
$\bullet$
 CMS Collaboration [CMS Collaboration],
  CMS-PAS-EXO-15-004.

\bibitem{Belyaev:2015hgo}
  A.~Belyaev, G.~Cacciapaglia, H.~Cai, T.~Flacke, A.~Parolini and H.~Serodio,
  arXiv:1512.07242 [hep-ph].

\bibitem{Bellazzini:2015nxw}
  B.~Bellazzini, R.~Franceschini, F.~Sala and J.~Serra,
  JHEP {\bf 1604} (2016) 072
  [arXiv:1512.05330 [hep-ph]].

 \bibitem{earlySp}
 Z.~y.~Duan, P.~S.~Rodrigues da Silva and F.~Sannino,
  Nucl.\ Phys.\ B {\bf 592} (2001) 371
  [hep-ph/0001303].

\bibitem{Katz:2005au}
  E.~Katz, A.~E.~Nelson and D.~G.~E.~Walker,
  JHEP {\bf 0508} (2005) 074
  [hep-ph/0504252].

\bibitem{Lodone:2008yy}
  P.~Lodone,
  JHEP {\bf 0812} (2008) 029
  [arXiv:0806.1472 [hep-ph]].

\bibitem{Gripaios:2009pe}
  B.~Gripaios, A.~Pomarol, F.~Riva and J.~Serra,
  JHEP {\bf 0904} (2009) 070
  [arXiv:0902.1483 [hep-ph]].

\bibitem{Cacciapaglia:2014uja}
  G.~Cacciapaglia and F.~Sannino,
  JHEP {\bf 1404} (2014) 111
  [arXiv:1402.0233 [hep-ph]].

\bibitem{Schmaltz:2010ac}
  M.~Schmaltz, D.~Stolarski and J.~Thaler,
  JHEP {\bf 1009} (2010) 018
  [arXiv:1006.1356 [hep-ph]].

\bibitem{Georgi:1984af}
  H.~Georgi and D.~B.~Kaplan,
  Phys.\ Lett.\ B {\bf 145} (1984) 216.

\bibitem{Dugan:1984hq}
  M.~J.~Dugan, H.~Georgi and D.~B.~Kaplan,
  Nucl.\ Phys.\ B {\bf 254} (1985) 299.

\bibitem{ArkaniHamed:2002qy}
  N.~Arkani-Hamed, A.~G.~Cohen, E.~Katz and A.~E.~Nelson,
  JHEP {\bf 0207} (2002) 034
  [hep-ph/0206021].

\bibitem{Vecchi:2013bja}
  L.~Vecchi,
  arXiv:1304.4579 [hep-ph].

\bibitem{Mrazek:2011iu}
  J.~Mrazek, A.~Pomarol, R.~Rattazzi, M.~Redi, J.~Serra and A.~Wulzer,
  Nucl.\ Phys.\ B {\bf 853} (2011) 1
  [arXiv:1105.5403 [hep-ph]].

\bibitem{Agashe:2006at}
  K.~Agashe, R.~Contino, L.~Da Rold and A.~Pomarol,
  Phys.\ Lett.\ B {\bf 641} (2006) 62
  [hep-ph/0605341].

\bibitem{Sikivie:1980hm}
  P.~Sikivie, L.~Susskind, M.~B.~Voloshin and V.~I.~Zakharov,
  Nucl.\ Phys.\ B {\bf 173} (1980) 189.

\bibitem{Coleman:1973jx}
  S.~R.~Coleman and E.~J.~Weinberg,
  Phys.\ Rev.\ D {\bf 7} (1973) 1888.

\bibitem{lattice}
  T.~DeGrand, Y.~Liu, E.~T.~Neil, Y.~Shamir and B.~Svetitsky,
  Phys.\ Rev.\ D {\bf 91} (2015) 11,  114502
  [arXiv:1501.05665 [hep-lat]].
%
$\bullet$
  T.~DeGrand, Y.~Liu, E.~T.~Neil, Y.~Shamir and B.~Svetitsky,
  PoS LATTICE {\bf 2014} (2014) 275
  [arXiv:1412.4851 [hep-lat]].

\bibitem{DeGrand:2015zxa}
  T.~DeGrand,
  doi:10.1103/RevModPhys.88.015001
  arXiv:1510.05018 [hep-ph].

\bibitem{higher}
  D.~K.~Hong, S.~D.~H.~Hsu and F.~Sannino,
  Phys.\ Lett.\ B {\bf 597} (2004) 89
  [hep-ph/0406200].
%
$\bullet$~D.~D.~Dietrich, F.~Sannino and K.~Tuominen,
  Phys.\ Rev.\ D {\bf 72} (2005) 055001
  [hep-ph/0505059].
  %
$\bullet$~D.~D.~Dietrich and F.~Sannino,
  Phys.\ Rev.\ D {\bf 75} (2007) 085018
  [hep-ph/0611341].

\bibitem{Golterman:2015zwa}
  M.~Golterman and Y.~Shamir,
  Phys.\ Rev.\ D {\bf 91} (2015) 9,  094506
  [arXiv:1502.00390 [hep-ph]].

\bibitem{Agashe:2004rs}
  K.~Agashe, R.~Contino and A.~Pomarol,
  Nucl.\ Phys.\ B {\bf 719} (2005) 165
  [hep-ph/0412089].

\bibitem{HiggsDiscovery}
  G.~Aad {\it et al.}  [ATLAS Collaboration],
  Phys.\ Lett.\ B {\bf 716} (2012) 1
  [arXiv:1207.7214 [hep-ex]].
%
$\bullet$~S.~Chatrchyan {\it et al.}  [CMS Collaboration],
  Phys.\ Lett.\ B {\bf 716} (2012) 30
  [arXiv:1207.7235 [hep-ex]].

\bibitem{WZW}
  J.~Wess and B.~Zumino,
  Phys.\ Lett.\ B {\bf 37} (1971) 95.
%
$\bullet$
  E.~Witten,
  Nucl.\ Phys.\ B {\bf 223}, 422 (1983).

\bibitem{Kaymakcalan:1983qq}
  O.~Kaymakcalan, S.~Rajeev and J.~Schechter,
  Phys.\ Rev.\ D {\bf 30} (1984) 594.

\bibitem{Georgi:1985nv}
  H.~Georgi and M.~Machacek,
  Nucl.\ Phys.\ B {\bf 262} (1985) 463.

  \bibitem{software}
  J.~Alwall {\it et al.},
  JHEP {\bf 1407} (2014) 079
  [arXiv:1405.0301 [hep-ph]].
  %
$\bullet$
 A.~Alloul, N.~D.~Christensen, C.~Degrande, C.~Duhr and B.~Fuks,
  Comput.\ Phys.\ Commun.\  {\bf 185} (2014) 2250
  [arXiv:1310.1921 [hep-ph]].

\bibitem{Frigerio:2012uc}
  M.~Frigerio, A.~Pomarol, F.~Riva and A.~Urbano,
  JHEP {\bf 1207} (2012) 015
  [arXiv:1204.2808 [hep-ph]].

\bibitem{Kim:2016jbz}
  M.~Kim, S.~J.~Lee and A.~Parolini,
  arXiv:1602.05590 [hep-ph].

\bibitem{Joseph:2009bq}
  A.~Joseph and S.~G.~Rajeev,
  Phys.\ Rev.\ D {\bf 80} (2009) 074009
  [arXiv:0905.2772 [hep-ph]].

\bibitem{Calibbi:2016ukt}
  L.~Calibbi, G.~Ferretti, D.~Milstead, C.~Petersson and R.~P\"ottgen,
  arXiv:1602.04821 [hep-ph].

\bibitem{oai:arXiv.org:0910.1789}
  B.~Gripaios,
  JHEP {\bf 1002} (2010) 045
  [arXiv:0910.1789 [hep-ph]].

\bibitem{Cacciapaglia:2015eqa}
  G.~Cacciapaglia, H.~Cai, A.~Deandrea, T.~Flacke, S.~J.~Lee and A.~Parolini,
  JHEP {\bf 1511} (2015) 201
  [arXiv:1507.02283 [hep-ph]].

\bibitem{Chatrchyan:2013wfa}
  S.~Chatrchyan {\it et al.} [CMS Collaboration],
  Phys.\ Rev.\ Lett.\  {\bf 112} (2014) no.17,  171801
  [arXiv:1312.2391 [hep-ex]].

\bibitem{DeGrand:2015yna}
  T.~DeGrand and Y.~Shamir,
  Phys.\ Rev.\ D {\bf 92} (2015) no.7,  075039
  [arXiv:1508.02581 [hep-ph]].

\bibitem{Pica:2016rmv}
  C.~Pica and F.~Sannino,
  arXiv:1604.02572 [hep-ph].

\bibitem{Kim:1984pt}
  J.~E.~Kim,
  Phys.\ Rev.\ D {\bf 31} (1985) 1733.

\bibitem{Farhi:1980xs}
  E.~Farhi and L.~Susskind,
  Phys.\ Rept.\  {\bf 74} (1981) 277.

\bibitem{Dashen:1969eg}
  R.~F.~Dashen,
  Phys.\ Rev.\  {\bf 183} (1969) 1245.

\bibitem{Dobrich:2015jyk}
  B.~Dobrich, J.~Jaeckel, F.~Kahlhoefer, A.~Ringwald and K.~Schmidt-Hoberg,
  JHEP {\bf 1602} (2016) 018
  [arXiv:1512.03069 [hep-ph]].

\bibitem{Anisimovsky:2004hr}
  V.~V.~Anisimovsky {\it et al.} [E949 Collaboration],
  Phys.\ Rev.\ Lett.\  {\bf 93} (2004) 031801
  [hep-ex/0403036].

\bibitem{Cadamuro:2011fd}
  D.~Cadamuro and J.~Redondo,
  JCAP {\bf 1202} (2012) 032
  [arXiv:1110.2895 [hep-ph]].

\bibitem{'tHooft:1976up}
  G.~'t Hooft,
  Phys.\ Rev.\ Lett.\  {\bf 37} (1976) 8.

\bibitem{Veneziano:1979ec}
  G.~Veneziano,
  Nucl.\ Phys.\ B {\bf 159} (1979) 213.

\bibitem{Witten:1979vv}
  E.~Witten,
  Nucl.\ Phys.\ B {\bf 156} (1979) 269.

\bibitem{Howe:2016mfq}
  K.~Howe, S.~Knapen and D.~J.~Robinson,
  arXiv:1603.08932 [hep-ph].

\bibitem{Cacciapaglia:2015vrx}
  G.~Cacciapaglia and A.~Parolini,
  Phys.\ Rev.\ D {\bf 93} (2016) no.7,  071701
  [arXiv:1511.05163 [hep-ph]].

\bibitem{'tHooft:1980xb}
  G.~'t Hooft,
  NATO Adv.\ Study Inst.\ Ser.\ B Phys.\  {\bf 59} (1980) pp.1.

\bibitem{CFTbounds}
  F.~Sannino,
  Phys.\ Rev.\ D {\bf 79} (2009) 096007
  [arXiv:0902.3494 [hep-ph]].
 %
 $\bullet$
T.~A.~Ryttov and F.~Sannino,
  Int.\ J.\ Mod.\ Phys.\ A {\bf 25} (2010) 4603
  [arXiv:0906.0307 [hep-ph]].

\bibitem{atheorem}
  J.~L.~Cardy,
  Phys.\ Lett.\ B {\bf 215} (1988) 749.
 %
 $\bullet$
  H.~Osborn,
  Phys.\ Lett.\ B {\bf 222} (1989) 97.
  %
 $\bullet$
  I.~Jack and H.~Osborn,
  Nucl.\ Phys.\ B {\bf 343} (1990) 647.

\end{thebibliography}
\end{document}